\documentclass[fleqn,usenatbib,useAMS]{mnras}

\usepackage{mathptmx}

\usepackage[british]{babel}
\usepackage[T1]{fontenc}
\usepackage{ae,aecompl}

\usepackage{graphicx}	
\usepackage{amsmath}	
\usepackage{amssymb}	
\usepackage{xspace}
\usepackage{verbatim}
\usepackage{color}
\usepackage{hyperref}

\newcommand{\galform}{{\sc galform}\xspace}
\newcommand{\mbh}{M_{\mathrm{BH}}}
\newcommand{\lbol}{L_{\mathrm{bol}}}

\newcommand{\ledd}{L_{\mathrm{Edd}}}

\newcommand{\angstrom}{\mbox{\normalfont\AA}}

\title[AGNs at the cosmic dawn]{AGNs at the cosmic dawn: predictions for future surveys from a $\Lambda$CDM cosmological model}

\author[A. J. Griffin et al.]{
Andrew J. Griffin,$^{1}$\thanks{E-mail: andrew.j.griffin@durham.ac.uk (AJG)}
Cedric G. Lacey,$^{1}$
Violeta Gonzalez-Perez,$^{2,3}$\newauthor
Claudia del P. Lagos,$^{4,5,6}$
Carlton M. Baugh,$^{1}$
Nikos Fanidakis$^{7,8}$
\\
$^{1}$Institute for Computational Cosmology, Department of Physics, University of Durham, South Road, Durham, DH1 3LE, UK\\
$^{2}$Institute of Cosmology and Gravitation, University of Portsmouth, Burnaby Road, Portsmouth PO1 3FX, UK\\
$^{3}$ Energy Lancaster, Lancaster University, Lancaster, LA1 4YB, UK\\
$^{4}$International Centre for Radio Astronomy Research (ICRAR), M468, University of Western Australia, 35 Stirling Hwy, Crawley, \\  WA 6009, Australia\\
$^{5}$ARC Centre of Excellence for All Sky Astrophysics in 3 Dimensions (ASTRO 3D)\\
$^{6}$Cosmic Dawn Center (DAWN), Copenhagen, Denmark\\
$^{7}$Max-Planck-Institute for Astronomy, K\"onigstugl 17, D-69117 Heidelberg, Germany\\
$^{8}$BASF, Carl-Bosch Strasse 38, 67056 Ludwigshafen, Germany
}

\date{Accepted XXX. Received YYY; in original form ZZZ}

\pubyear{2020}

\begin{document}
\label{firstpage}
\pagerange{\pageref{firstpage}--\pageref{lastpage}}
\maketitle

\begin{abstract}
Telescopes to be launched over the next decade-and-a-half, such as JWST, EUCLID, ATHENA and Lynx, promise to revolutionise the study of the high redshift Universe and greatly advance our understanding of the early stages of galaxy formation.
We use a model that follows the evolution of the masses and spins of supermassive black holes (SMBHs) within a semi-analytic model of galaxy formation to make predictions for the Active Galactic Nucleus (AGN) luminosity function at $z\geq7$
in the broadband filters of JWST and EUCLID at near-infrared wavelengths, and ATHENA and Lynx at X-ray energies. 
The predictions of our model are relatively insensitive to the choice of seed black hole mass, except at the lowest luminosities ($\lbol<10^{43}\mathrm{ergs^{-1}}$) and the highest redshifts ($z>10$).
We predict that surveys with these different telescopes will select somewhat different 
samples of SMBHs, with EUCLID unveiling the most massive, highest accretion rate SMBHs, Lynx the least massive, lowest
accretion rate SMBHs, and JWST and ATHENA covering objects inbetween. 
At $z=7$, we predict that typical detectable SMBHs will have masses, $M_{\mathrm{BH}}\sim10^{5-8}M_{\odot}$, and Eddington normalised mass accretion rates, 
$\dot{M}/\dot{M}_{\mathrm{Edd}}\sim0.6-2$. The SMBHs will be hosted by galaxies of stellar mass $M_{\star}\sim10^{8-10}M_{\odot}$, and dark matter haloes of mass $M_{\mathrm{halo}}\sim10^{11-12}M_{\odot}$. We predict that the detectable SMBHs at $z=10$ will have slightly smaller black holes, accreting at slightly higher Eddington normalised mass accretion rates, in slightly lower mass host galaxies compared to those at $z=7$, and reside in haloes of mass $M_{\mathrm{halo}}\sim10^{10-11}M_{\odot}$.

\end{abstract}




\section{Introduction}

Recent advances in observational capabilities have allowed us to investigate AGNs in the early Universe more thoroughly than ever before. At optical wavelengths, the Sloan Digital Sky Survey \citep[SDSS,][]{york00} initiated the hunt for quasars out to redshift $z \sim 6$ \citep{fan01a,fan03,fan04,jiang09}. Detections at $z \sim 6$ of fainter quasars have been made by the Canada-France High-z Quasar Survey \citep[CFHQS,][]{willott10}, and a quasar has been detected at $z=7.1$ in the United Kingdom Infrared Deep Sky Survey \citep[UKIDSS,][]{lawrence07} by \cite{mortlock11}. Currently, the highest redshift quasar known is at $z=7.64$, as discovered at optical/near-infrared wavelengths by mining three large area surveys \citep{banados18a}, and the same object has also been observed at X-ray wavelengths using Chandra \citep{banados18b}. Recent radio observations using the Giant Metrewave Radio Telescope have also been able to detect AGNs at high redshift, such as a radio galaxy at $z=5.72$ \citep{saxena18}. 

AGNs are believed to play an important role in galaxy formation at low redshift, as they are seen to produce huge X-ray cavities in the hot intracluster gas in galaxy clusters \citep[e.g.][]{forman05,david11,cavagnolo11}, and AGN feedback is included in theoretical models of galaxy formation to shut off gas cooling in massive haloes and star formation in the largest galaxies \citep[e.g.][]{dimatteo05,croton06a,bower06}, in order to reproduce the bright end of the galaxy luminosity function. AGNs may also play an important role in galaxy formation at higher redshift, where large-scale outflows driven by AGNs are observed e.g. at $z \sim 2$ \citep{harrison12}, and at $z \sim 6$ \citep{maiolino12,cicone15}. X-ray observations have also indicated that faint QSOs may play an important role in reionising the Universe \citep{giallongo15,onoue17,ricci17}.

At $z \sim 6$, AGNs have been discovered with estimated black hole masses over a billion solar masses \citep[e.g.][]{willott10b,derosa11,venemans13,wu15}. How these SMBHs could grow to such large masses in such a short time is a puzzle. SMBHs grow from seed black holes, which could form from remnants of a first generation of (Population III) stars, or from gas clouds that form supermassive stars that eventually collapse to form a black hole, or from dense star clusters that collapse via stellar dynamical processes \citep[e.g.][]{volonteri10}. These seeds are expected to be of mass $M_{\mathrm{seed}} = 10-10^{5} M_{\odot}$ depending on the formation mechanism, with the remnants of Population III stars forming light ($\sim 10-100 M_{\odot}$) seeds, gas cloud collapse forming heavy ($\sim 10^{4-5} M_{\odot}$) seeds, and star cluster collapse forming seeds of intermediate ($\sim 10^3 M_{\odot}$) mass \citep{volonteri10}. SMBHs can then grow either by accretion of gas or by merging with other SMBHs. To form the observed high redshift SMBHs by gas accretion, these seeds require sustained accretion near the Eddington rate for several hundred Myr, which may be interrupted by feedback effects.


Fortunately, the next decade-and-a-half promise to be exciting for observing the high redshift Universe. The launch of the James Webb Space Telescope (JWST) in 2021 will pave the way for an increased understanding of the $z>7$ Universe \citep[e.g.][]{gardner06,kalirai18}. JWST, with its 6.5m diameter mirror, will make observations from the optical to mid-infrared (0.6 $\mu$m to 30 $\mu$m) to probe the earliest galaxies and the stars contained within them. EUCLID, also due for launch in 2021, with a 1.2m diameter mirror, is primarily a cosmology mission with the aim of constraining dark energy, but the surveys it will conduct at optical and near-IR wavelengths (0.5-2 $\mu$m) will also be useful for detecting high-redshift quasars \citep{laureijs11,barnett19}. While JWST and EUCLID will probe similar wavelength ranges, the specifications of the missions are different. The sensitivity of JWST is better, but EUCLID will survey much larger areas of sky, which will lead to different samples of AGNs being detected by these two missions, as they will sample AGNs with different luminosities and space densities.

The Advanced Telescope for High-ENergy Astrophysics (ATHENA) \citep{nandra13}, scheduled for launch in 2031, will observe the high-redshift Universe at X-ray energies (0.5-10 keV). The Lynx X-ray observatory \citep{lynx18}, which has a proposed launch date in 2035, will also observe the distant Universe at similar energies (0.2-10 keV). The science objectives of both missions include determining the nature of SMBH seeds and investigating the influence of SMBHs on the formation of the first galaxies. The two missions have different capabilities: ATHENA has a larger field of view and larger effective area (which leads to better instrumental sensitivity) at 6 keV, but a worse angular resolution and lower effective area at 1 keV, compared to Lynx. The improved angular resolution of Lynx results in better sensitivity in practice, as sources that would be affected by source confusion when observed by ATHENA would be unaffected if observed by Lynx. Therefore, the two telescopes will detect different luminosity objects. 

We are now entering an era in which the properties of SMBHs in the high redshift Universe ($z>7$) during the first billion years of its evolution can be robustly probed. By comparing observations with simulations, we can test theoretical models of galaxy formation, and by comparing to the high redshift Universe, we can test these theoretical models in a regime that up to now is poorly constrained. 

In this paper, we present predictions for the AGN population at $z \geq 7$ for comparison with observations from JWST, EUCLID, ATHENA, and Lynx, using the model for SMBH and AGN evolution presented in \cite{griffin19} (hereafter Paper I), which includes a self-consistent treatment of SMBH spin, to predict AGN luminosities. This paper is one of a series of papers exploring SMBH and AGN properties within a physical galaxy formation model based on the $\Lambda$CDM model of structure formation. Paper I presented the model for the evolution of SMBH and AGN within the \cite{baugh19} recalibration of the \cite{lacey16} \galform semi-analytical model of galaxy formation, showing a comparison of the predicted SMBH and AGN properties to observations for $0 \leq z \leq 6$. Here, we extend the predictions of this model to $z \geq 7$.

Other theoretical models have also made predictions for the evolution of SMBHs and AGNs through cosmic time, such as semi-analytic models \citep[e.g.][]{lagos08,marulli08,bonoli09,fani12,hirschmann12,menci13,neistein14,enoki14,shirakata19}, hydrodynamical simulations \citep[e.g.][]{hirschmann14,sijacki15,rosasguevara16,weinberger18}, and more empirical models \citep[e.g.][]{saxena17,weigel17}. Predictions have also been made for $z>7$ using empirical models \citep[e.g.][]{aird13,barnett19} and semi-analytic models \citep[e.g.][]{ricarte18b}. In this paper, we are making predictions for AGNs at $z \geq 7$ from a semi-analytic galaxy formation model, which includes more channels of SMBH growth than \cite{ricarte18b}. A few predictions from our model have also previously been shown in \cite{amarantidis19}, in which AGN luminosity functions from several different theoretical models are compared.

This paper is structured as follows. 
In Section \ref{sec:model} we outline the model used. In Section \ref{sec:mbh} we present predictions for black hole properties, and in Section \ref{sec:lbol_high_z} we present predictions for AGN luminosity functions for $z \geq 7$. In 
Section \ref{sec:predictions} we present predictions for AGNs detectable by future surveys using JWST, EUCLID, ATHENA and Lynx, and
in Section \ref{sec:conclusions} we give our conclusions.


\section{Method}
\label{sec:model}

In this paper, we analyse the properties of SMBHs and AGNs within the \galform semi-analytic model of galaxy formation. We briefly outline the galaxy formation model, and the modelling of SMBHs and AGNs, which follow Paper I, apart from one change described below. 

\subsection{The \galform galaxy formation model}

In this paper, we present predictions using the same galaxy formation model as Paper I, which is the \cite{baugh19} recalibration of the \cite{lacey16} \galform model. \galform is a semi-analytic model of galaxy formation, which was introduced in \cite{cole00}. In \galform, galaxies form in dark matter haloes, with the evolution of the dark matter haloes described by halo merger trees. For a recent full description of the model, see \cite{lacey16}. In the model used here, the halo merger trees are extracted from a cosmological dark matter N-body simulation \citep{helly03}. The baryonic exchange between different components (e.g. stars, hot halo gas, cold disc gas, black hole) is modelled by a set of coupled differential equations. Physical processes modelled in \galform include 
i) the merging of dark matter haloes,
ii) shock heating and radiative cooling of gas in haloes,
iii) collapse of cooled gas onto a rotationally supported disc, 
iv) a two-phase interstellar medium for the cold gas with star formation from molecular gas,
v) feedback from photoionisation, supernovae, and AGNs,
vi) the chemical evolution of gas and stars, 
vii) galaxies merging in haloes due to dynamical friction,
viii) bar instabilties in galaxy discs, 
ix) the evolution of stellar populations, and 
x) the extinction and reprocessing of stellar radiation by dust.
The analytical prescriptions for these processes include a number of free parameters, which are calibrated on a range of observational constraints on galaxy properties. 

\galform has undergone continual development, with various \galform models now in existence, \citep[e.g.][]{gonzalezperez14,lacey16}. This paper uses the \cite{baugh19} recalibration of the \cite{lacey16} \galform model for the Planck cosmology. This recalibrated model was presented for use with P-Millennium dark matter merger trees \citep{baugh19}. P-Millennium is a high resolution dark matter simulation using the \cite{planck14} cosmology, with a box of side $800$Mpc and a halo mass resolution of $2.12 \times 10^9 h^{-1} M_{\odot}$ (corresponding to 20 particles). The \cite{lacey16} model matches to a wide range of observational data, both in terms of wavelength (from far-UV luminosity functions to sub-mm number counts), and in terms of redshift, matching a large range of observations from $z \sim 0$ to $z \sim 6$.

\subsection{SMBHs and AGNs in \galform}

SMBHs start out as seed black holes, which we model by adding a seed black hole of mass $M_{\mathrm{seed}}$ to each dark matter halo. Unless otherwise stated, the value of $M_{\mathrm{seed}}$ adopted is $10h^{-1} M_{\odot}$. SMBHs can then grow via three channels: (i) starbursts triggered by mergers or disc instabilities, which can drive gas to the galaxy centre to be made available for accretion onto the SMBH (ii) `hot halo' accretion in which gas quiescently accretes from the hot gas atmosphere in the largest haloes and (iii) mergers between SMBHs. Unlike some other models, the gas accretion rate is not assumed to be Eddington-limited.

Building on \cite{fani11}, in Paper I a model for the evolution of SMBH spin was presented, in which SMBH spin evolves via accretion of gas, or by merging with another SMBH. The SMBH/AGN model involves several free parameters, for which we use the same values as in Paper I. In Paper I, we generally adopted values from previous studies, with two free parameters ($\eta_{\mathrm{Edd}}$, which controls the suppression of luminosity for super-Eddington accretion rates, and $f_{\mathrm{q}}$ which determines the lifetimes of the AGN episodes) calibrated on the observed AGN bolometric luminosity function for $0 \leq z \leq 6$. 

In the starburst mode, we assume that the SMBH accretion rate is constant over a time:

\begin{equation}
    t_{\mathrm{acc}} = f_{\mathrm{q}} t_{\mathrm{bulge}} \label{eq:tacc_stb}
\end{equation}

\noindent
where $t_{\mathrm{bulge}}$ is the dynamical timescale of the bulge. In Paper I, we gave the equations for bolometric radiative AGN luminosities in different accretion regimes: i) an Advection Dominated Accretion Flow (ADAF) state accreting via a physically thick, optically thin disc \citep{narayanyi94}, ii) a thin disc state accreting via a physically thin, optically thick disc \citep{shakurasunyaev73}, and iii) a super-Eddington state accreting via a slim disc \citep{abramowicz88}. We use these same equations in this paper, except for a slightly modified expression for the luminosity in the super-Eddington regime, where for Eddington normalised mass accretion rates $\dot{m} > \eta_{\mathrm{Edd}} (0.1/ \epsilon(a))$, the bolometric luminosity is now given by:

\begin{equation}
    L_{\mathrm{bol}} = \eta_{\mathrm{Edd}} \Big( 1+ \ln \Big( \frac{\dot{m}}{\eta_{\mathrm{Edd}}} \frac{\epsilon(a)}{0.1} \Big) \Big) \ledd , \label{eq:lbol}
\end{equation}

\noindent
where $\epsilon(a)$ is the spin-dependent radiative accretion efficiency for a thin accretion disc, $a$ is the dimensionless spin parameter, $\eta_{\mathrm{Edd}}$ is a free parameter, $\dot{m} = \dot{M} / \dot{M}_{\mathrm{Edd}}$ is the Eddington normalised mass accretion rate, and $\ledd$ is the Eddington luminosity. The Eddington luminosity is given by:

\begin{equation}
 \ledd = 1.26 \times 10^{46} \, \Big( \frac{\mbh}{10^8 M_{\odot}} \Big) \, \rm{ergs^{-1}},
\end{equation}

\noindent
and we define the Eddington mass accretion rate by:

\begin{equation}
 \dot{M}_{\mathrm{Edd}} = \frac{\ledd}{0.1 c^2}. \label{eq:mdot_edd}
\end{equation}

We use a nominal accretion efficiency, $\epsilon=0.1$ in equation (\ref{eq:mdot_edd}), so that the Eddington normalised mass accretion rate does not depend on the spin \citep[this is a commonly used convention, cf.][]{yn14}. 
The slight modification to the bolometric luminosities for the super-Eddington regime compared to Paper I ensures that the luminosities vary continuously in the transition from thin disc to super-Eddington accretion rates.

The model calculates luminosities at near-IR to X-ray wavelengths from the bolometric luminosities using the template SED in \cite{marconi04}. This SED is empirical, where the ratio of luminosities at 2500$\angstrom$ and 2 keV is a function of bolometric luminosity, such that the optical emission dominates at high bolometric luminosities, and the X-ray emission dominates at low bolometric luminosities. This AGN SED was used in Paper I, where we showed that this model gives good agreement with observed optical/UV and X-ray AGN luminosity functions for $0 \leq z \leq 5$. In this paper, we use the same SED model to extend our model predictions to higher redshifts. 

We investigated the possible effect of using a more physical AGN SED model by comparing the \cite{marconi04} SED to the \cite{netzer19} SED model. In the latter, the optical/UV emission is modelled assuming a standard thin accretion disc that emits as a blackbody at each radius. This yields an optical/UV spectrum dependent on the SMBH mass, mass accretion rate and spin. The X-ray luminosities are then related to the optical luminosities using an empirical power law relation. When we compare this to the \cite{marconi04} SED model, we find that at $1400 \angstrom$, the bolometric correction factors are in very good agreement for $10^{42} \mathrm{ergs^{-1}} < L_{1400} < 10^{46} \mathrm{ergs^{-1}}$. For $L_{1400} > 10^{46} \mathrm{ergs^{-1}}$ the two bolometric correction factors are in poorer agreement, but this would only have a small effect on our predictions as there are very few objects at these high optical luminosities in our simulaton box. At hard X-ray energies, the two bolometric correction factors differ by a factor of about 2. This means that if we were to use the \cite{netzer19} bolometric correction for our predictions, this would result in objects having higher X-ray luminosities, increasing the number of objects predicted to be detected at X-ray energies.

We plan in a future study to make AGN predictions using a more physical AGN SED model, such as that of \cite{kubotadone18}, in which the X-ray emission originates from hot and warm coronae, which are dependent on the SMBH mass and accretion rate. We defer any further discussion of the effect of changing the AGN SED model to this future study. 


AGNs are understood to be surrounded by a gas and dust torus, which absorbs radiation from the AGN, the absorbed radiation then being re-emitted at longer (IR) wavelengths. To model this obscuration effect, we use several empirical relations for the `visible fraction' which is the fraction of AGNs that are not obscured by a torus at a given luminosity, redshift, and wavelength (see Section 3.3 of Paper I).

\subsection{AGN model variants}
\label{sec:variants}

In Paper I, we showed that the fiducial model overpredicts the rest-frame $1500 \angstrom$ and soft X-ray AGN luminosity functions at $z=6$, and so alongside predictions for the fiducial model, we presented two alternative models with slight modifications that provide a better fit to these AGN luminosity functions. The three models and the visible fractions used are as follows (see Paper I for more details):

\begin{enumerate}

\item First, our fiducial model which uses the `low-z modified Hopkins' (LZMH) visible fraction, which has a functional form that is based on the obscuration model used in \cite{hopkinsrh07}, but with different coefficients. The visible fraction for rest-frame $1500\angstrom$ is:

\begin{equation}
 f_{\mathrm{vis, 1500, LZMH}} = 0.15 \Big( \frac{\lbol}{10^{46} \rm{ergs^{-1}}}\Big)^{-0.1}, \label{eq:LZMH_opt}
\end{equation}

\noindent
where $\lbol$ is the bolometric luminosity. For the rest-frame soft X-ray band (0.5-2 keV), the visible fraction is:

\begin{equation}
 f_{\mathrm{vis, SX, LZMH}} = 0.4 \Big( \frac{\lbol}{10^{46} \rm{ergs^{-1}}}\Big)^{0.1}. \label{eq:LZMH_SX}
\end{equation}

As in Paper I, in this paper we assume that there is no obscuration for the hard X-ray band (2-10 keV). The coefficients for the visible fraction were derived in Paper I by constructing an observational bolometric luminosity function from the observational optical/UV, soft X-ray, and hard X-ray luminosity functions. The luminosities at these different wavelengths were converted to bolometric luminosities using the \cite{marconi04} SED, and the number densities were converted to total number densities using the assumed visible fractions. The coefficients of the visible fractions are then chosen by eye to give the smallest scatter in the resultant bolometric luminosity function.

\item The second of the models uses the `$z=6$ modified Hopkins' (Z6MH) visible fraction, which is:

\begin{equation}
 f_{\mathrm{vis, Z6MH}} = 0.04. \label{eq:Z6MH}
\end{equation}

This value was obtained by selecting coefficients in the power-law expressions for the visible fraction that result in the best agreement with the rest-frame $1500 \angstrom$ and rest-frame soft X-ray AGN luminosity functions at $z=6$.

\item The third of the models used in this paper is the `low accretion efficiency' model, which uses the LZMH visible fraction, but the fraction of mass accreted onto an SMBH in each starburst is lower. This was implemented in the model by decreasing the value of $f_{\mathrm{BH}}$, which represents the fraction of the mass of stars formed in a starburst that is accreted on to the SMBH in the form of gas. The modified value is $0.002$, compared to $0.005$ in the fiducial model. The luminosity suppression for super-Eddington sources was also varied, with the parameter $\eta_{\mathrm{Edd}}$ being increased to 16, compared to a value of 4 in the fiducial model. As for the previous variant of the model, these values were chosen to give agreement with the observed rest-frame $1500 \angstrom$ and rest-frame soft X-ray luminosity functions at $z=6$. This low accretion efficiency model predicts fewer objects than the fiducial model. 

\end{enumerate}

\section{Black hole mass function and accretion rates}
\label{sec:mbh}

\begin{figure}
\centering
\includegraphics[width=\linewidth]{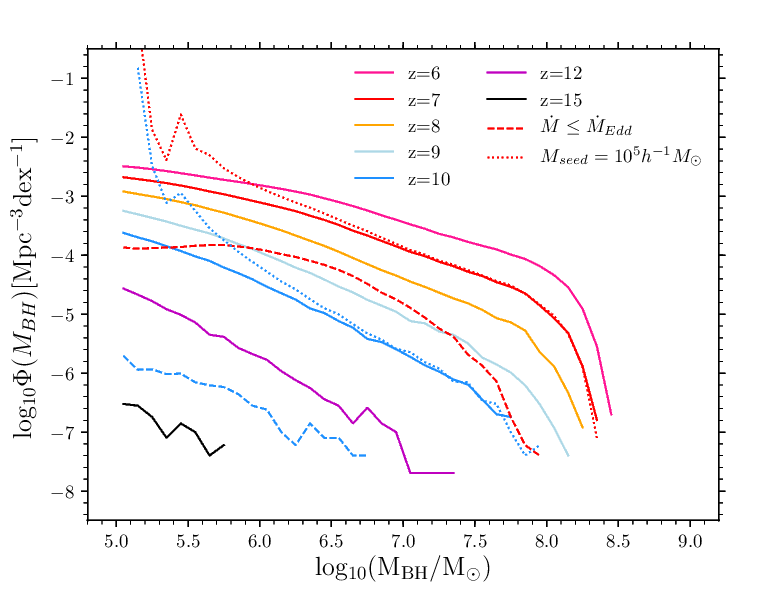}
\caption{The black hole mass function in the fiducial model for $z=6$ (pink solid line), $z=7$ (red solid line), $z=8$ (yellow solid line), $z=9$ (light blue solid line), $z=10$ (blue solid line), $z=12$ (purple solid line), and $z=15$ (black solid line). We also show the black hole mass functions when the gas accretion rate is not allowed to exceed the Eddington mass accretion rate for $z=7$ (red dashed line) and $z=10$ (blue dashed line). We show the black hole mass function for a seed mass of $10^{5} h^{-1} M_{\odot}$, for $z=7$ (red dotted line) and at $z=10$ (blue dotted line).}
\label{fig:BHMF_highZ} 
\end{figure}

In Figure \ref{fig:BHMF_highZ} we show the black hole mass function, $\Phi(M_{\mathrm{BH}})$, predicted by the model over the range $6<z<15$. We define $\Phi(X) = dn / d \log X$ throughout this paper, where $n$ is comoving number density, except for luminosity functions expressed in AB magntitudes, $M_{AB}$, where $\Phi(M) = dn / dM_{AB}$. Black holes build up in the model as a result of galaxies forming in dark matter haloes, which build up hierarchically. In the model, for our simulation volume of $(800 \mathrm{Mpc})^3$, some SMBHs of mass $10^8 M_{\odot}$ have already formed by $z=9$, but at $z=6$ there are no SMBHs with masses above $\mbh = 3 \times 10^8 M_{\odot}$. This appears to be in conflict with observations of extremely massive SMBHs at $z=6$ \citep[e.g.][]{willott10b,derosa11,venemans13,wu15}, which find estimated masses up to $\sim (0.3-1) \times 10^{10} M_{\odot}$. The lack of these objects in this simulation may be because high-redshift surveys probe larger volumes than the volume of the simulation box in this work (e.g. the total survey volume for \cite{banados18a} is of order 10 Gpc$^3$ compared to the volume of $0.5$ Gpc$^3$ for this simulation), and so are able to detect rarer objects \citep[e.g.][]{amarantidis19}. There are also uncertainties in the observational black hole mass estimates due to the use of observationally calibrated relations to determine black hole masses from observed emission line widths and luminosities. These errors are a mixture of random (these relations have an intrinsic scatter of a factor of about 3 \citep[e.g.][]{vestergaard06}), and systematic (these relations are only constrained for certain luminosity ranges in the local Universe).  

We also show in Figure \ref{fig:BHMF_highZ} the predicted black hole mass function for the case in which gas accretion onto SMBHs in the model is not allowed to exceed the Eddington mass accretion rate (i.e. $\dot{M} \leq \dot{M}_{\mathrm{Edd}}$)\footnote{Note that this is not the same as the low accretion efficiency model, where we reduce the fraction of gas accreting onto SMBHs in starbursts.}. In our standard model, SMBHs are allowed to accrete mass at super-Eddington accretion rates, and it can be seen that restricting SMBH accretion rates to the Eddington rate results in many fewer high-redshift SMBHs. At $z=7$, restricting SMBH accretion in this way causes the number of SMBHs to decrease by about 1 dex at $\mbh = 10^{6-7} M_{\odot}$, and by about 1.5 dex at $\mbh = 10^{5} M_{\odot}$ and 2.5 dex at $\mbh = 10^{8} M_{\odot}$. At $z=10$, the effect of restricting SMBH growth is even more significant, with the number density of SMBHs decreasing by about 2 dex at $\mbh = 10^{5-7} M_{\odot}$. This shows the importance of super-Eddington accretion in building up high-redshift SMBHs in our model.

We also show the black hole mass function at $z=7$ and $z=10$ when a seed mass, $M_{\mathrm{seed}} = 10^5 h^{-1} M_{\odot}$ is adopted, instead of $M_{\mathrm{seed}} = 10 h^{-1} M_{\odot}$ as in the fiducial model. At both of these redshifts, there are a large number of black holes around the seed mass for this case, but at higher masses the black hole mass function converges to the same value as in the fiducial model. This shows how the SMBH masses are relatively unaffected by the choice of seed black hole mass for sufficiently high SMBH mass provided that the gas accretion rate is not Eddington limited.

\begin{figure}
\centering
\includegraphics[width=\linewidth]{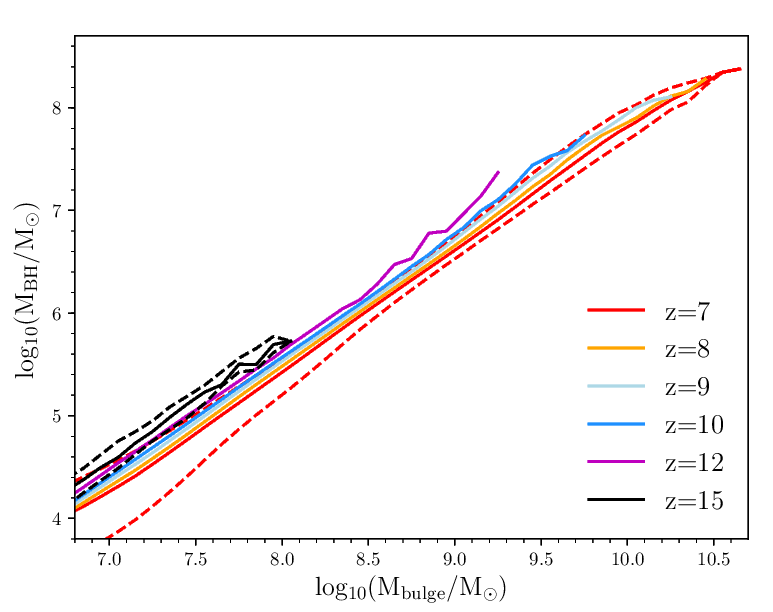}
\caption{The predicted SMBH mass versus bulge stellar mass relation at different redshifts, as indicated by the legend. The lines represent the median SMBH mass for each bin in bulge stellar mass. We also show the 10-90 percentiles of these distributions at $z=7$ (red dashed lines) and at $z=15$ (black dashed lines).}
\label{fig:SMBH_Mbulge_relation_highZ} 
\end{figure}

In Figure \ref{fig:SMBH_Mbulge_relation_highZ}, we present the predicted evolution of the SMBH mass versus bulge stellar mass relation for $7 \leq z \leq 15$. This relation evolves only weakly with redshift, with the SMBH mass at a given bulge mass increasing by a factor $\sim 2$ from $z=7$ to $z=15$. This is a continuation of the trend seen in Paper I for $0 \leq z \leq 6$. This trend occurs in the model because at higher redshifts, bulges grow mostly by starbursts which grow both the SMBH and the bulge, whereas at lower redshifts, starbursts are less prevalent, and stars are transferred from the disc to the bulge in merger or disc instability events, without growing the SMBH. The evolution of the SMBH mass versus total stellar mass is generally similar to the evolution of the SMBH mass versus bulge stellar mass, with SMBHs having a higher mass at a given total stellar mass at higher redshift. However, as shown in Figure 5 of Paper I, at lower total stellar masses, the slope of the relation is steeper compared to higher masses. At these low masses, the evolution with redshift is slightly stronger.

\begin{figure}
\centering
\includegraphics[width=\linewidth]{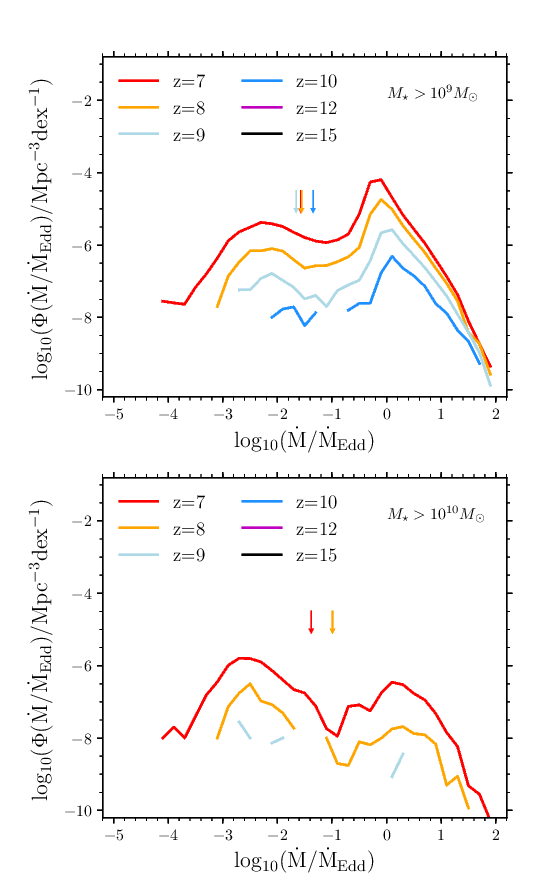}
\caption{The number density of objects as a function of Eddington normalised mass accretion rate, $\dot{M}/ \dot{M}_{\mathrm{Edd}}$, at $z=7$ (red), $z=8$ (yellow), $z=9$ (light blue), $z=10$ (dark blue), $z=12$ (purple), and $z=15$ (black). We show the median of each distribution as a downward pointing arrow. Only SMBHs residing in galaxies with stellar masses above $M_{\star} = 10^9 M_{\odot}$ are shown in the upper panel, whereas this stellar mass threshold is $M_{\star} = 10^{10} M_{\odot}$ for the lower panel. No curves are plotted for $z \geq 12$ in the upper panel and for $z \geq 10$ in the lower panel because either there are too few objects to plot a curve or there are no objects in our simulation volume above the cuts in stellar mass that we are applying.}
\label{fig:MMEdd_distribution_highZ} 
\end{figure}

In Figure \ref{fig:MMEdd_distribution_highZ} we show the number of objects as a function of Eddington normalised mass accretion rate ($\dot{M}/ \dot{M}_{\mathrm{Edd}}$) predicted by the model at $7 \leq z \leq 15$, for SMBHs residing in galaxies with stellar masses above $10^9 M_{\odot}$ or $10^{10} M_{\odot}$. At each redshift, the distribution is bimodal, with peaks at $\dot{M}/ \dot{M}_{\mathrm{Edd}} \sim 0.001$, and $\dot{M}/ \dot{M}_{\mathrm{Edd}} \sim 1$. The peak at $\dot{M}/ \dot{M}_{\mathrm{Edd}} \sim 1$ is produced by AGNs fuelled by starbursts triggered by disc instabilities. The value of $\dot{M}/ \dot{M}_{\mathrm{Edd}}$ at this peak increases slightly with redshift, which is a result of galaxy bulges having a smaller dynamical timescale at higher redshift, which results in shorter accretion timescales (cf. equation (\ref{eq:tacc_stb})). Galaxies have lower masses at higher redshift, and so the mass of gas transferred in each disc instability episode is typically smaller at higher redshift, and SMBHs are smaller at higher redshift. The former decreases $\dot{M}/ \dot{M}_{\mathrm{Edd}}$, while the latter increases $\dot{M}/ \dot{M}_{\mathrm{Edd}}$, and these effects almost cancel out. 

The peak at $\dot{M}/ \dot{M}_{\mathrm{Edd}} \sim 0.001$ is produced by AGNs fuelled by hot halo accretion. There is also a minor contribution from AGNs fuelled by starbursts triggered by mergers with $\dot{M}/ \dot{M}_{\mathrm{Edd}}$ values in the range 0.1-1. The peak at $\dot{M}/ \dot{M}_{\mathrm{Edd}} \sim 1$ has more objects when the stellar mass cut is $10^9 M_{\odot}$, but the peak at $\dot{M}/ \dot{M}_{\mathrm{Edd}} \sim 0.001$ has more objects when the stellar mass cut is $10^{10} M_{\odot}$. This is because AGNs fuelled by starbursts triggered by disc instabilities reside in lower stellar mass galaxies than AGNs fuelled by hot halo accretion. We allow SMBHs to accrete above the Eddington mass accretion rate in our model, and in this figure we see that there are objects that accrete at super-Eddington rates, but none above $\dot{M}/ \dot{M}_{\mathrm{Edd}}=100$. 

\begin{figure}
\centering
\includegraphics[width=\linewidth]{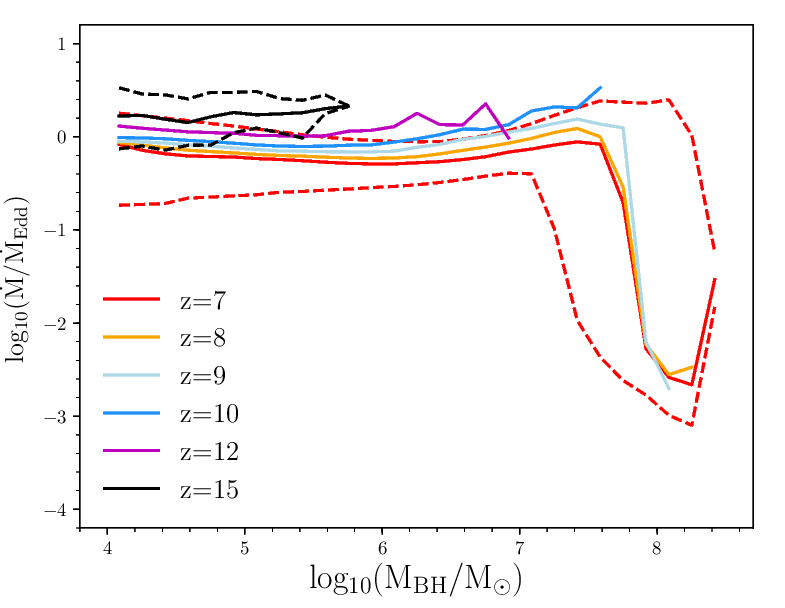}
\caption{The Eddington normalised mass accretion rate ($\dot{M}/ \dot{M}_{\mathrm{Edd}}$) versus SMBH mass relation, at different redshifts, as indicated by the legend. The lines represent the median $\dot{M}/ \dot{M}_{\mathrm{Edd}}$ for each bin in SMBH mass. We also show the 10-90 percentiles of these distributions at $z=7$ (red dashed lines) and at $z=15$ (black dashed lines).}
\label{fig:mdot_SMBH_highZ} 
\end{figure}

In Figure \ref{fig:mdot_SMBH_highZ}, we show the predicted evolution of the Eddington normalised mass accretion rate ($\dot{M} / \dot{M}_{\mathrm{Edd}}$) versus SMBH mass. For $\mbh \lesssim 10^{7.5} M_{\odot}$, this relation is generally flat, but for $\mbh \gtrsim 10^{7.5} M_{\odot}$, the average $\dot{M} / \dot{M}_{\mathrm{Edd}}$ decreases dramatically. This is because at the highest masses, SMBHs are more likely to be fuelled by the hot halo mode, and because this involves quiescent accretion onto large SMBHs, the $\dot{M} / \dot{M}_{\mathrm{Edd}}$ values are lower. The average $\dot{M} / \dot{M}_{\mathrm{Edd}}$ at a given SMBH mass generally increases slightly with redshift. The trend of SMBHs having slightly higher typical $\dot{M} / \dot{M}_{\mathrm{Edd}}$ values at higher redshifts is also seen in Figure \ref{fig:MMEdd_distribution_highZ}.


\section{Evolution of AGN luminosities for z > 7}
\label{sec:lbol_high_z}

\begin{figure*}
\centering
\includegraphics[width=\linewidth]{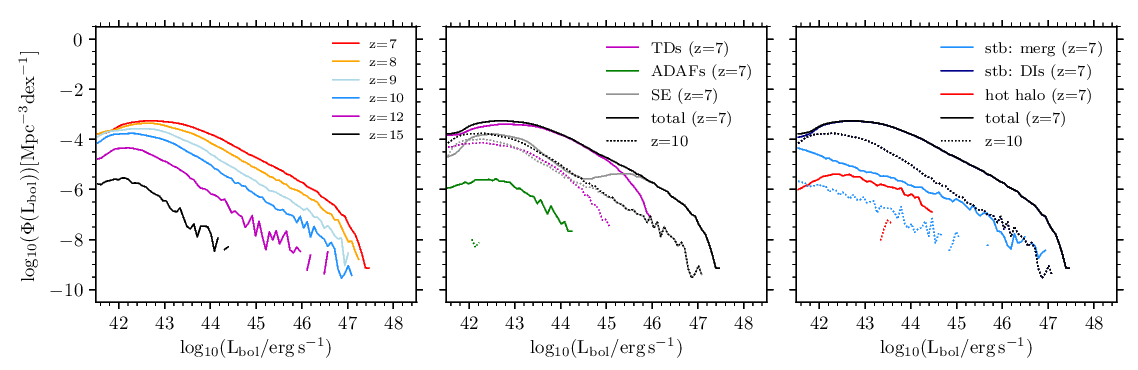}
\caption{The predicted AGN bolometric luminosity function for the fiducial model at high redshift. \emph{Left panel}: The evolution of the bolometric
luminosity function for $z=7$ (black), $z=8$ (red), $z=9$ (yellow), $z=10$ (green), $z=12$ (light blue), $z=15$ (purple).
The turnover at low luminosity is due to the halo mass resolution.
\emph{Middle panel}: The AGN bolometric luminosity function at $z=7$ (solid lines) and $z=10$ (dotted lines), showing the total luminosity function (black), and the contribution from ADAFs (green), thin discs (purple)
and super-Eddington objects (grey). \emph{Right panel}: The AGN bolometric luminosity function at $z=7$ (solid lines) and $z=10$ (dotted lines), showing the total luminosity function (black) and the contribution from objects fuelled by the hot halo mode (red),
starbursts triggered by mergers (light blue) and starbursts triggered by disc instabilities (dark blue). Note that the dark blue lines are under the black lines.}
\label{fig:bol_LF_high_z_2} 
\end{figure*}

In the left panel of Figure \ref{fig:bol_LF_high_z_2}, we show the evolution of the AGN bolometric luminosity function  
for the fiducial model for $7 \leq z \leq 15$.
As the redshift increases, both the number of objects and the luminosities decrease.
By $z \approx 12$, there are almost no objects brighter than $\lbol \sim 10^{46} \mathrm{ergs^{-1}}$
in our simulated volume of (800Mpc)$^3$. 

We have investigated the effects of halo mass resolution on our predictions. In Figure \ref{fig:Resolution_bol_lf} we show the bolometric luminosity function for the standard model (with a halo mass resolution of $2.12 \times 10^{9} h^{-1} M_{\odot}$) alongside the model with a halo mass resolution of $10^{10} h^{-1} M_{\odot}$. This comparison shows that the turnover in the bolometric luminosity function at low luminosity is due to halo mass resolution. The bolometric luminosity functions are converged for $\lbol > 10^{43} \mathrm{ergs^{-1}}$. 

In Figure \ref{fig:Seed_bol}, we explore the effect of varying the black hole seed mass on the AGN bolometric luminosity function. We find that the AGN bolometric luminosity function is not sensitive to the choice of seed black hole mass for values in the range $M_{\mathrm{seed}}=(10-10^5) h^{-1} M_{\odot}$ for $\lbol > 10^{42} \mathrm{ergs^{-1}}$ at $z=7$, and for $\lbol > 10^{43} \mathrm{ergs^{-1}}$ at $z=12$. For luminosities below this, the seed mass does affect the predictions.

In the middle panel of Figure \ref{fig:bol_LF_high_z_2} we split the AGN luminosity function at $z=7$ and $z=10$ into the contributions
from ADAFs, thin discs and super-Eddington objects. 
Paper I showed that at $z=0$, the contribution from ADAFs dominates the predicted AGN luminosity function at low luminosities ($\lbol < 10^{44} \mathrm{ergs^{-1}}$), while the contribution from thin discs dominates at intermediate luminosities ($10^{44} \mathrm{ergs^{-1}} < \lbol < 10^{46} \mathrm{ergs^{-1}}$) and the contribution from super-Eddington objects dominates at high luminosities ($\lbol > 10^{46} \mathrm{ergs^{-1}}$). As redshift increases, the contribution from ADAFs decreases, and the contribution from thin discs dominates at low luminosities, while the contribution from super-Eddington objects continues to dominate at high luminosities. This trend continues with increasing redshift, so that by $z=10$, the contribution from ADAFs is extremely small. At low luminosities ($\lbol < 10^{45} \mathrm{ergs^{-1}}$), the thin disc contribution and the contribution from super-Eddington objects are then approximately equal, while
at higher luminosities super-Eddington objects dominate. This implies that most of the QSOs (with $\lbol > 10^{45} \mathrm{ergs}^{-1}$)
that will be detectable by surveys conducted by future telescopes at $z=10$ should be accreting above the Eddington rate. This prediction is not straightforward to test, as determining Eddington ratios requires estimations of black hole masses. Black hole masses can be estimated from measurements of emission line widths, or black hole masses and mass accretion rates can be determined by fitting theoretical SED models to multi-wavelength data \citep[e.g.][]{kubotadone18}. The black hole masses estimated using either of these methods will have some model dependencies.

In the right panel of Figure \ref{fig:bol_LF_high_z_2} we split the AGN luminosity function at $z=7$ and $z=10$ by gas fuelling mode, into hot halo mode,
and starbursts triggered by galaxy mergers or disc instabilities. The dominant contributor at all luminosities at both $z=7$ and $z=10$ is starbursts 
triggered by disc instabilities, 
so we predict that future high-redshift surveys will detect AGNs fuelled by this mechanism. This prediction contrasts with some other theoretical models. Some hydrodynamical simulations predict that gas may be driven into the centres of galaxies by high density cold streams for accretion onto the SMBH \citep[e.g.][]{khandai12,dimatteo17}, while some other semi-analytical models simply assume that merger triggered starbursts dominate SMBH growth at high-redshift \citep[e.g.][]{ricarte18a}.

\begin{figure}
\centering
\includegraphics[width=\linewidth]{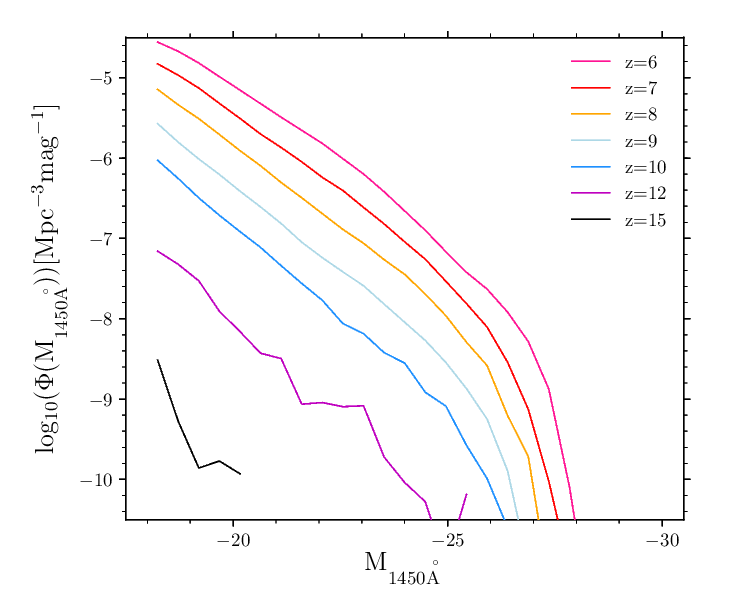}
\caption{The rest-frame $1450 \angstrom$ AGN luminosity function predicted by the model at different redshifts, as indicated by the legend.}
\label{fig:optical_1500_z6_z9} 
\end{figure}

In Figure \ref{fig:optical_1500_z6_z9}, we present the rest-frame $1450 \angstrom$ AGN luminosity function, $\Phi(M_{1450}) = dn / dM_{1450}$, predicted by the model for $6 \leq z \leq 15$, where $M_{1450}$ is the absolute AB magnitude at $1450 \angstrom$. Similarly to the AGN bolometric luminosity function, the rest-frame $1450 \angstrom$ AGN luminosity function decreases to lower number densities and lower luminosities with increasing redshift as a result of hierarchical structure formation. For comparison with observational studies \citep[e.g.][]{jiang16}, and empirical models \citep[e.g.][]{barnett19}, we calculate the density evolution parameter $k$, where $k$ is given by:

\begin{equation}
    \Phi(z, < M_{1450}) = \Phi(z_{0}, <M_{1450}) 10^{k(z-z_{0})},
\end{equation}

\noindent
and $\Phi(z, < M_{1450})$ is the cumulative number density of AGN for $M_{1450}$ brighter than the given value. For $5 \leq z \leq 6$, the model evolves less strongly than the observations, as discussed in Section \ref{sec:variants}, where we present two variants of the model to address this. For $6 \leq z \leq 7$, our fiducial model predicts $k=-0.56$ for $M_{1450} < -25.3$ and $k=-1.08$ for $M_{1450} < -26.7$, whereas \cite{jiang16} find $k=-0.60 \pm 0.36$ and $k=-0.92 \pm 0.41$ for these same optical magnitudes. Our model is therefore consistent with the evolution measured by \cite{jiang16} at these optical magnitudes. For $6 \leq z \leq 9$, our fiducial model predicts $k=-0.50$ for $M_{1450}<-24$, and $k=-0.80$ for $M_{1450}<-26$. For the same redshift interval, \cite{barnett19} adopt $k=-0.72$ (their standard model), and $k=-0.92$ (their more steeply declining model). Our model is thus consistent with the evolution adopted in \cite{barnett19} for $M_{1450}<-26$, but not for $M_{1450}<-24$. Our model variant which uses the Z6MH visible fraction evolves with very similar $k$ values to the fiducial model, and our modified accretion efficiency model variant evolves with slightly more negative values of $k$. This model is still consistent with \cite{jiang16}, and is consistent with \cite{barnett19} for $M_{1450}<-24$ and $M_{1450}<-26$.

\begin{figure}
\centering
\includegraphics[width=\linewidth]{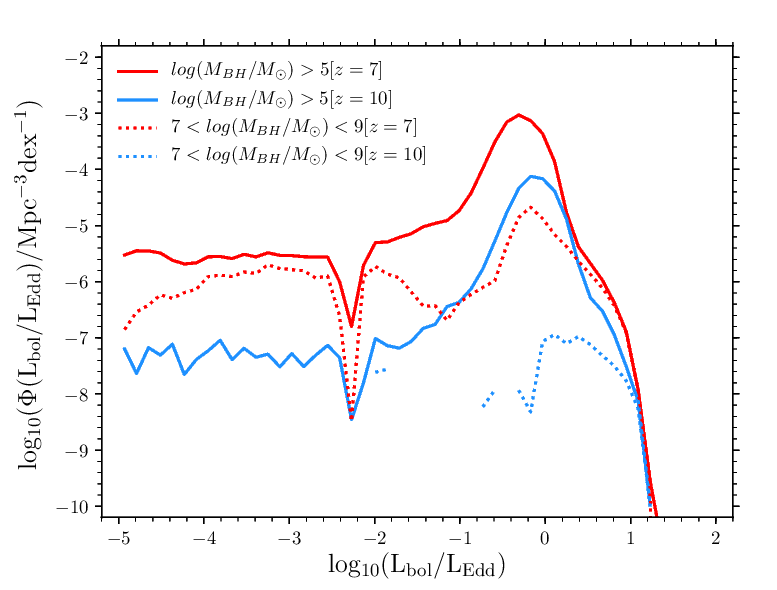}
\caption{The number density of objects as a function of Eddington normalised luminosity, $L / \ledd$, predicted by the model at $z=7$ (red) and $z=10$ (blue), for SMBHs with mass $\mbh > 10^5 M_{\odot}$ (solid lines), and for SMBHs with mass $10^{7} M_{\odot} < \mbh < 10^{9} M_{\odot}$ (dotted lines).}
\label{fig:LLEdd_distribution_highZ} 
\end{figure}

In Figure \ref{fig:LLEdd_distribution_highZ}, we present the number of objects as a function of $L / \ledd$ predicted by the model for $z=7$ and $z=10$ for black holes with $\mbh > 10^5 M_{\odot}$. The distributions are flat for $L / \ledd < 0.1$, and peak at $L / \ledd \sim 1$. The $L / \ledd$ value of the peak of the distribution slightly increases with redshift. There are no objects with $L / \ledd > 10$ in our simulated volume at these redshifts, which is a result of there being no objects with $\dot{M} / \dot{M}_{\mathrm{Edd}} > 100$ combined with our luminosity suppression for super-Eddington sources (cf. equation (\ref{eq:lbol})). The sharp dip around $L / \ledd = 0.01$ arises from the thin disc to ADAF transition not being continuous in luminosity.

We also show in Figure \ref{fig:LLEdd_distribution_highZ} the distribution of $L / \ledd$ predicted by the model for $10^{7} M_{\odot} < \mbh < 10^{9} M_{\odot}$, alongside the distribution for $\mbh > 10^5 M_{\odot}$. At $z=7$, black holes in these two mass ranges have similar distributions of $L / \ledd$ values, while for $z=10$, the number of black holes for $10^{7} M_{\odot} < \mbh < 10^{9} M_{\odot}$ in our simulation is too small to draw any strong conclusion on the form of this distribution.

\begin{figure*}
\centering
\includegraphics[width=\linewidth]{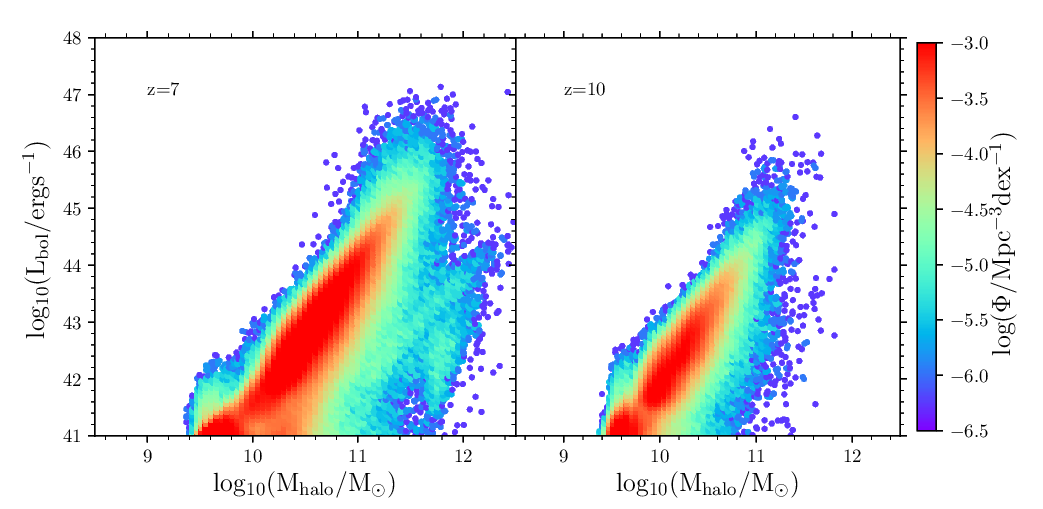}
\caption{A scatter plot of AGN bolometric luminosity versus host halo mass for AGNs at $z=7$ (left panel) and $z=10$ (right panel). The colour indicates the number density of objects.}
\label{fig:Lbol_Mhalo_z7_z10} 
\end{figure*}

In Figure \ref{fig:Lbol_Mhalo_z7_z10}, we present the AGN bolometric luminosity versus host halo mass for objects in the model, colour-coded by the number density of objects. The objects mostly follow a relation between bolometric luminosity and halo mass, although there are some objects offset from this relation to higher halo masses at $z=7$, but not at $z=10$. The objects on the main relation are fuelled by starbursts triggered by disc instabilities, whereas the objects offset from the main relation at higher halo masses are fuelled by hot halo mode accretion. The brightest model AGNs are not hosted by the most massive haloes at $z=7$, but at $z=10$ the brightest model AGNs are hosted by the most massive haloes, as a result of there being no objects fuelled by the hot halo mode by $z=10$.


\begin{table*}
\caption{The sensitivities and solid angles covered by the possible surveys by JWST, EUCLID, ATHENA and Lynx. For ATHENA and Lynx, the survey area is assumed is that of a single field of view, whereas for JWST and EUCLID the survey area is assumed to be that of multiple fields of view. The integration time is the total for a survey in that band. For ATHENA and Lynx, the flux limits used are the estimated confusion limits. These flux limits, $f_{\nu}$, can be related to apparent AB magnitudes by: $m_{\mathrm{AB}}= 31.40 - 2.5 \log_{10}(f_{\nu}/\mathrm{nJy})$.}
\begin{tabular}{ |c|c|c|c|c|c| } 
\hline
Instrument & Filter & $\lambda (\rm{\mu m})$ or E(keV) & Flux Limit & Survey Area  & Assumed total \\ 
& & & & & integration time (ks) \\
\hline
JWST NIRCam & F200W & $1.7-2.3~\rm{\mu m}$ & 9.1 nJy & $9680$ $\, \rm{arcmin^{2}}$ (1000 FoVs) & 10000 \\
 & F444W & $3.8-5.1~\rm{\mu m}$ & 23.6 nJy & $9680$ $\, \rm{arcmin^{2}}$ (1000 FoVs) & 10000 \\
\hline
EUCLID (Deep & H & $1.5-2~\rm{\mu m}$ & 145 nJy & $40 \, \rm{deg^{2}} \, (70 \, FoVs)$ & $\sim 13000$ \\
 Survey) & & & & & \\
\hline
EUCLID (Wide & H & $1.5-2~\rm{\mu m}$ & 912 nJy & $15000 \, \rm{deg^{2}} \, (26000 \, FoVs)$ & $\sim 120000$ \\
 Survey) & & & & & \\
\hline
ATHENA WFI & Soft X-ray & $0.5 - 2~ \mathrm{keV}$ & $2.4 \times 10^{-17} \mathrm{erg} \, \mathrm{cm^{-2} s^{-1}}$ & $1600 \, \rm{arcmin^{2}}$ (FoV) & 450 \\
 & Hard X-ray & $2 - 10~ \mathrm{keV}$ & $1.6 \times 10^{-16} \mathrm{erg} \, \mathrm{cm^{-2} s^{-1}}$ & $1600 \, \rm{arcmin^{2}}$ (FoV) & 450 \\
\hline
Lynx & Soft X-ray & $0.5 - 2~ \mathrm{keV}$ & $7.8 \times 10^{-20} \mathrm{erg} \, \mathrm{cm^{-2} s^{-1}}$ & $360 \, \rm{arcmin^{2}}$ (FoV) & 15000 \\
 & Hard X-ray & $2 - 10~ \mathrm{keV}$ & $1.0 \times 10^{-19} \mathrm{erg} \, \mathrm{cm^{-2} s^{-1}}$ & $360 \, \rm{arcmin^{2}}$ (FoV) & 15000 \\
\hline
\end{tabular}
\label{tab:sensitivities}
\end{table*}

\begin{figure*}
\centering
\includegraphics[width=.8\linewidth]{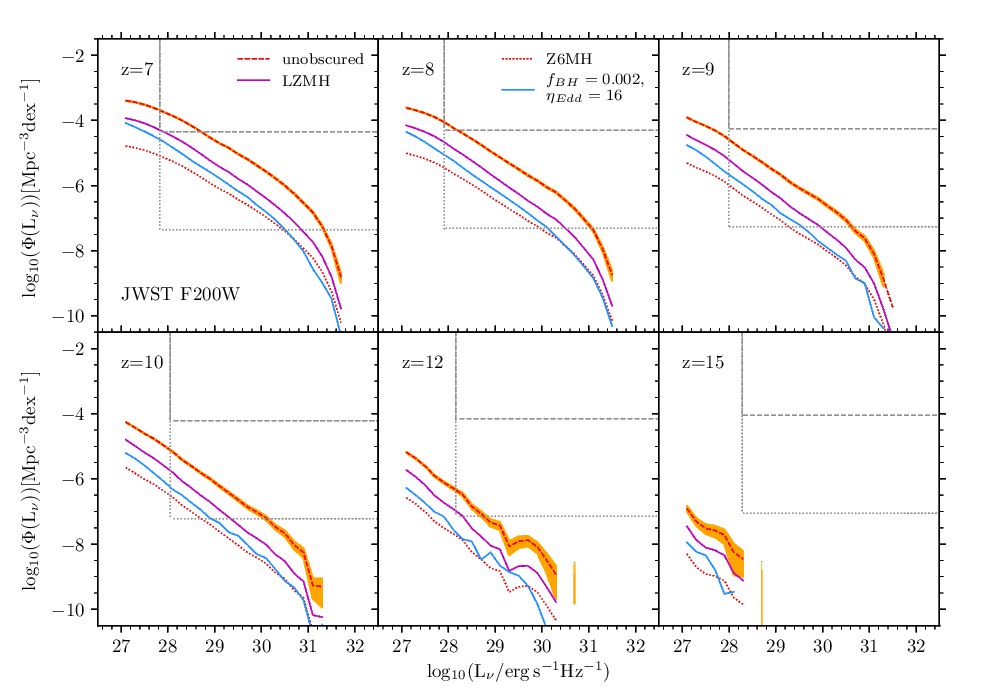}
\caption{Predictions for the AGN luminosity function in the observer frame JWST NIRCam F200W (2.0$\mu$m) band.
We show the luminosity function for the fiducial model without 
obscuration (red dashed) with Poisson errors (orange shading), the fiducial model with the 
`low z modified Hopkins' (LZMH) visible fraction (magenta solid), the fiducial model with the `$z=6$ modified Hopkins' (Z6MH) visible fraction (red dotted),
and the low accretion efficiency model which uses the `low z modified Hopkins' visible fraction (blue solid).
The horizontal lines indicate the number density limit resulting from a survey area of one field of view (dashed), and the number density limit resulting from 1000 of these fields of view (dotted). The vertical lines show the luminosity limit resulting from the flux limit. The assumed flux limits and survey areas are given in Table \ref{tab:sensitivities}. Detectable objects are above and to the right of these lines. These luminosities can be converted into absolute AB magnitudes via $M_{\mathrm{AB}}= 51.59 - 2.5 \log (L_{\nu}/ \mathrm{erg \, s^{-1} Hz^{-1}})$.}
\label{fig:JWST_NIR_F200W_lf} 
\end{figure*}


\section{Predictions for high redshift surveys with future telescopes}
\label{sec:predictions}

We next employ our model to make predictions for the detection of AGNs at $z \geq 7$ with the future telescopes described in the Introduction. We use luminosity functions predicted by the model in the different wavelength or energy bands of these telescopes to predict the number of AGNs that should be detectable by surveys with these telescopes. We also describe the typical properties of the SMBHs detectable by the different telescopes. The survey parameters that we assume for JWST \footnote{\url{https://jwst-docs.stsci.edu/display/JTI/NIRCam+Sensitivity}}, EUCLID\footnote{\url{https://www.euclid-ec.org/?page_id=2581}}, ATHENA\footnote{\url{https://www.cosmos.esa.int/documents/400752/507693/Athena_SciRd_iss1v5.pdf}}, and Lynx\footnote{\url{https://wwwastro.msfc.nasa.gov/lynx/docs/LynxInterimReport.pdf}} are summarised in Table \ref{tab:sensitivities}.

The number of AGNs detectable in a survey depends on both the flux limit and the survey area. The former affects the ability to detect low luminosity sources and the latter affects the number density of objects down to which one can probe. In practice, the luminosity of the host galaxy also sets a limit on identifying AGNs in deep surveys, but we do not take this into account here.

From the predicted flux limits of the surveys, luminosity limits can be derived using $L = 4 \pi d_{\mathrm{L}}^{2} f$ for calculating broadband luminosities (ATHENA and Lynx) and $L_{\nu} = 4 \pi d_{\mathrm{L}}^{2} f_{\nu} / (1+z)$ for calculating a luminosity per unit frequency (EUCLID and JWST). Here, $f$ is the flux, $f_{\nu}$ is the flux per unit frequency and $d_{\mathrm{L}}$ is the luminosity distance to the source, $L$ is the luminosity in the rest-frame band or wavelength corresponding to the observed band or wavelength, and $L_{\nu}$ is the luminosity per unit frequency in the rest frame corresponding to the observed wavelength and redshift. We use these expressions to calculate luminosity limits (vertical lines) in Figures \ref{fig:JWST_NIR_F200W_lf} to \ref{fig:Athena_Lynx_HX}. 

The luminosities shown in Figures \ref{fig:JWST_NIR_F200W_lf} to \ref{fig:Athena_Lynx_HX} have been k-corrected to a fixed band in the observer frame. Our template SED for this calculation is that of \cite{marconi04}, for which the ratio of X-ray to optical luminosity varies with bolometric luminosity. To calculate the luminosity in each band we input the bolometric luminosity and the redshift and then integrate the SED over frequency multiplied by the appropriate response function for the filter redshifted into the rest frame of the source. There is a one-to-one relation between bolometric luminosity and luminosity in a particular band.

The number density limit for a survey can be calculated via the following method. The number of objects per log flux per unit solid angle per unit redshift is given by:

\begin{equation}
 \frac{\mathrm{d}^{3}N}{\mathrm{d(log}f_{\nu}) \mathrm{d}z \mathrm{d} \Omega} = \frac{\mathrm{d}^2 N}{\mathrm{d(log}L_{\nu}) \mathrm{d}V} \frac{\mathrm{d}^2 V }{\mathrm{d}z \mathrm{d} \Omega},
\end{equation}

\noindent
where $V$ is the comoving volume, $\mathrm{d}^2 N / \mathrm{d(log} L_{\nu}) \mathrm{d}V$ is the luminosity function in comoving units, and $\mathrm{d}^2 V / \mathrm{d}z \mathrm{d} \Omega$ is the comoving volume per unit solid angle per unit redshift. We define $\Phi(X) = \mathrm{d}^2 N / \mathrm{d(log} X) \mathrm{d}V$ so the luminosity function can be written as $\Phi (L_{\nu})$. For there to be an average of at least one object detectable in the survey per log flux per unit redshift, we therefore have the condition:

\begin{equation}
 \frac{\mathrm{d}^2 N}{\mathrm{dlog}L_{\nu} \mathrm{d}V} \geq \frac{1}{\frac{\mathrm{d}^2 V}{\mathrm{d}z \mathrm{d} \Omega}\Delta\Omega},
\end{equation}

\noindent
where $\Delta\Omega$ is the solid angle of sky covered by the survey. This condition allows us to construct the number density limits (horizontal lines) in Figures \ref{fig:JWST_NIR_F200W_lf} to \ref{fig:Athena_Lynx_HX}. Note that this limit is almost independent of redshift over the range $7 \leq z \leq 15$, as also seen for the JWST predictions of \cite{cowley18} for galaxies. The flux limits and survey areas adopted for the predictions for different telescopes are given in Table \ref{tab:sensitivities}. These limits then allow us to predict the number of objects detectable by each survey, for the three different model variants, as given in Table \ref{tab:number_of_objects}, and the properties of these objects, for the fiducial model, as given in Tables \ref{tab:nature_of_objects_median_z7}, and \ref{tab:nature_of_objects_median_z10}. 

In general, the flux limit determines the lower luminosity limit of objects that can be detected, whereas the survey area determines the upper luminosity limit of objects that can be detected. The different flux limits and survey areas of the surveys conducted by the different telescopes therefore provide detections of different populations of AGNs. 

We do not present predictions for $z \geq 7$ AGN detections by JWST or EUCLID at observer frame optical wavelengths. This is because at these redshifts, neutral hydrogen in the Inter-Galactic Medium absorbs all radiation at wavelengths less than that of Lyman-$\alpha$ emission in the rest-frame of the AGN ($1216 \angstrom$) \citep{gunn65,fan06}.

\subsection{Near-IR surveys with JWST and EUCLID}

JWST, planned for launch in 2021, will observe at wavelengths of 0.6-29 $\mu$m. It will have instruments for both imaging and spectroscopy,
including the NIRCam for optical to near-infrared imaging (0.7-5 $\mu$m) and MIRI for mid-infrared imaging (5-29 $\mu$m). We present predictions for two different NIRCam bands. We do not make predictions for MIRI,
because our AGN model does not currently include emission from the dust torus, which would be necessary for modelling AGN emission in the mid-infrared.
Figure \ref{fig:JWST_NIR_F200W_lf} shows predicted AGN luminosity functions in the observer frame F200W (2.0$\mu$m) band.
We also find that in the observer frame F444W (4.4$\mu$m) band, the predicted luminosity functions are similar to the observer frame F200W band. We present predictions for a survey composed of 1000 fields of view, each with a $10^4$s integration time, giving a total integration time of $10^7$s in each band. Figure \ref{fig:JWST_NIR_F200W_lf} shows that the effect of obscuration causes the predicted number of AGNs to be 0.04-0.2 of the predicted number of objects if obscuration is not taken into account. The effect of low accretion efficiency causes the predicted number of objects to be about 0.4 times lower than in the fiducial model if we are assuming the LZMH obscuration model. We predict that on average, $<1$ AGN per unit $z$ per field of view will be detectable by JWST for a $10^4$s integration, once we allow for obscuration.

We give the predicted number of objects for each survey in Table \ref{tab:number_of_objects}. For JWST we are assuming a survey of 1000 fields of view, each with a $10^4$s integration time per band. We predict that 
$90-500$ AGNs (depending on which of the three models is used) will be observed at $z=7$ in the F200W band and $60-300$ in the F444W band. We predict that more objects will be detectable in the F200W band because the assumed flux limit for the F200W band is lower than for the F444W band, which translates into a lower limit for the bolometric luminosity and higher number density. Predictions for the number of objects detectable at $z=9$, $z=10$ and $z=12$ are given in Table \ref{tab:number_of_objects}.

From the flux limits in these bands, limits in bolometric luminosity can be calculated. At $z=7$, we predict that JWST will detect AGNs with bolometric luminosities in the range $(6 \times 10^{43}-3 \times 10^{46}) \, \mathrm{ergs^{-1}}$ (F200W), and $(1 \times 10^{44}-4 \times 10^{46}) \, \mathrm{ergs^{-1}}$ (F444W). For the assumed survey parameters, we predict that JWST will be able to detect AGNs out to $z=9$ for all the near-IR bands, with F200W being more favourable for detecting $z>7$ AGNs than F444W. For F200W, we predict that about 60-90 times fewer AGNs will be detectable at $z=10$ than at $z=7$. Considering even higher redshift objects, for $z>10$ we predict that detection with JWST will become more difficult, as AGNs become extremely rare as well as very faint. 

We also explored whether a wide JWST survey composed of 1000 fields of view (as in Table \ref{tab:sensitivities}) or a deep survey composed of one field of view for an integration time 1000 times longer (10Ms) would detect more objects. We found that the deep survey would detect more AGNs ($300-2000$) than the wide survey ($90-500$) in the F200W band at $z=7$, although in practice the number of AGNs identifiable in the deep survey might be reduced by contamination by light from their host galaxies. 

\begin{figure*}
\centering
\includegraphics[width=.8\linewidth]{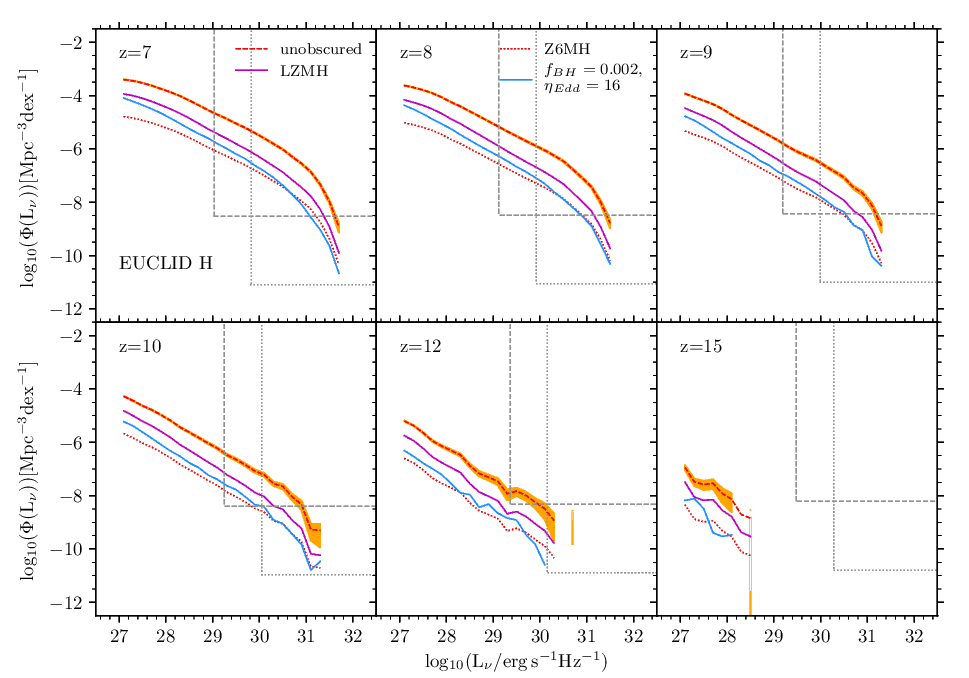}
\caption{Predictions for the AGN luminosity
function in the observer frame EUCLID H (1.5-2 $\mu$m) band. 
The dashed lines represent the sensitivity and survey volume limits of 
the EUCLID Deep survey and the dotted lines represent the sensitivity and survey volume limits of the EUCLID Wide survey.}
\label{fig:EUCLID_H_distant_lf} 
\end{figure*}

EUCLID, due for launch in 2021, will use its visible and near-IR coverage (0.55-2 $\mu$m) of galaxies to probe the nature of dark energy, but these same surveys will also allow detections of high-redshift AGNs.  EUCLID will conduct two surveys: a Wide Survey covering 15000 deg$^2$ of sky
and a Deep Survey covering 40 deg$^2$ in three fields. The mission lifetime of EUCLID will be 6.25 years. The surveys will be conducted in four bands - one visible (VIS) and
three near-IR (Y,J,H). 
We show predictions for the EUCLID H (1.5-2$\mu$m) band in Figure
\ref{fig:EUCLID_H_distant_lf}.
We show the sensitivity and survey volume limits for both the Deep and Wide surveys. The two surveys are seen to be quite complementary for detecting high redshift AGNs at different luminosities.

At $z=7$, we predict that the EUCLID H band will detect AGNs with bolometric luminosities $\lbol = (7 \times 10^{44}- 1 \times 10^{47}) \, \mathrm{ergs^{-1}}$ for the Deep Survey, and with $\lbol = (4 \times 10^{45}-3 \times 10^{47}) \, \mathrm{ergs^{-1}}$ for the Wide Survey. We therefore predict that the two EUCLID surveys and surveys by JWST will sample different parts of the AGN luminosity function.

At $z=7$, we predict that $100-600$ AGNs will be detectable in the EUCLID Deep survey using the H band (depending on the model), whereas for the Wide survey at $z=7$, we predict that $(8-30) \times 10^3$ AGNs will be detectable. At $z=10$, we predict $1-5$ AGNs will be detectable in the Deep survey, and $70-300$ in the Wide survey. AGNs are detectable in the H band at $z=10$ because the peak of the observed SED is at $1.3 \mu$m, close to the H band wavelength. It may be that such observations will reveal that the AGN SED shape at high redshift is different to the \cite{marconi04} SED used in this work.

According to our model, it will be impossible to detect very high redshift ($z=15$) objects with EUCLID, so such investigation may have to wait until
surveys after EUCLID. This is because despite the survey area being sufficiently large to probe down to the required number densities, the sensitivity of EUCLID is not sufficient to detect these low luminosity AGNs.

The alternative models featuring a lower visible fraction or lower accretion efficiency predict fewer AGNs than the fiducial model, 
so observations using EUCLID and JWST may be able to differentiate between these models as well as constraining the form of the AGN SED and thus 
provide better understanding of the high redshift AGN population.

\begin{figure*}
\centering
\includegraphics[width=.8\linewidth]{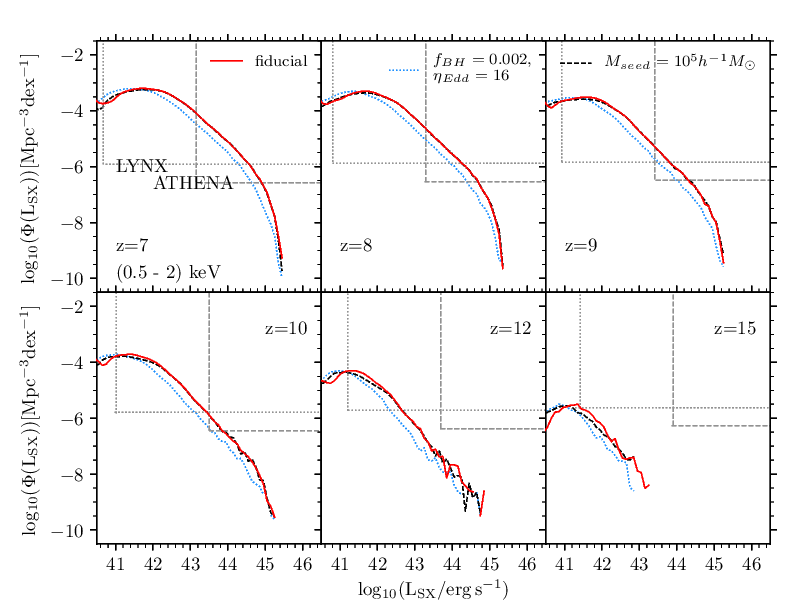}
\caption{Predictions for AGN luminosity functions in the observer frame soft X-ray band. Shown are the fiducial model (red solid line), the low accretion efficiency model (blue dotted line), and the fiducial model with seed black hole mass $10^{5} h^{-1} M_{\odot}$ (black dashed line). We also show the ATHENA (dashed) and Lynx (dotted) luminosity and number density limits (vertical and horizontal lines) for a single field of view and integration down to the estimated confusion limit, as in Table \ref{tab:sensitivities}.}
\label{fig:Athena_Lynx_SX} 
\end{figure*}

\begin{figure*}
\centering
\includegraphics[width=.8\linewidth]{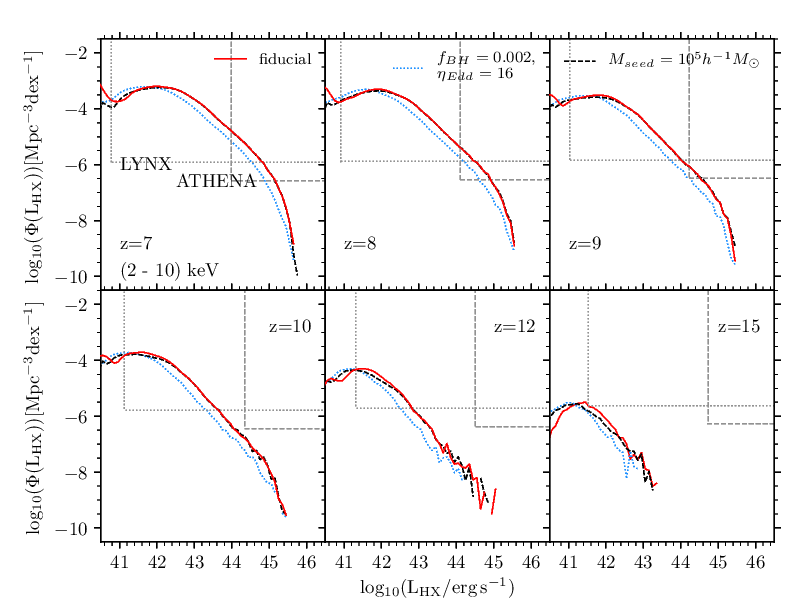}
\caption{As for Figure \ref{fig:Athena_Lynx_SX}, but for the observer frame hard X-ray band.}
\label{fig:Athena_Lynx_HX} 
\end{figure*}

\subsection{X-ray surveys with ATHENA and Lynx}

Due for launch in 2031, ATHENA will make observations at 0.5-10 keV using two instruments: the X-ray Integral Field Unit (X-IFU) for high resolution spectroscopy
and the Wide Field Imager (WFI) with a large field of view for surveys \citep{nandra13}. The Lynx X-ray observatory, with a proposed launch date of 2035, will make observations at 0.2-10 keV. Due to the effects of source confusion, Lynx will be able to probe down to lower luminosities than ATHENA as a result of its much better angular resolution.

We have calculated the sensitivity limits due to source confusion for ATHENA and Lynx. Source confusion occurs when multiple sources are separated by angles less than the angular resolution of the telescope and so appear merged together in images. To derive the confusion limits for ATHENA and Lynx, we use the commonly used \cite{condon74} `source density criterion', to obtain the cumulative number count per solid angle at the confusion limit ($N(>f_{\mathrm{conf}})$), for a given beam solid angle, $\Omega_{\mathrm{beam}}$, and number of beams per source $\mathcal{N}_{\mathrm{beam}}$: 

\begin{equation}
    N(>f_{\mathrm{conf}}) = 1 / \mathcal{N}_{\mathrm{beam}} \Omega_{\mathrm{beam}}, \label{eq:n_f_conf}
\end{equation}

\noindent
where the beam solid angle is related to the full width half maximum (FWHM) telescope beam width, $\theta_{\mathrm{FWHM}}$, by $\Omega_{beam}= \pi \theta_{\mathrm{FWHM}}^2 / (4 (\gamma-1) \ln 2)$ for a Gaussian beam profile, where $\gamma$ is the slope of the power law relating differential number count and flux, given by:

\begin{equation}
    \frac{d^2 N}{df d\Omega} \propto f^{-\gamma}.
\end{equation}

We use $\mathcal{N}_{\mathrm{beam}}=30$. Having calculated the cumulative number count at the confusion limit from equation (\ref{eq:n_f_conf}), we can obtain the flux at the confusion limit by using a model that relates the cumulative number counts to the flux. For this, we use the \cite{lehmer12} empirical model, which is a fit to the number counts measured using Chandra assuming a double power law fit for the AGN contribution, and single power law fits for the galaxy and stellar contributions. For the Lynx sensitivities, we are extrapolating the \cite{lehmer12} model to 100-1000 times lower fluxes than observed by Chandra. For ATHENA, $\theta_{\mathrm{FWHM}}=5 \, \mathrm{arcsec}$, whereas for Lynx, $\theta_{\mathrm{FWHM}}=0.5 \, \mathrm{arcsec}$. The $\gamma$ values that we use are slopes of the differential number counts from \cite{lehmer12} at the estimated confusion limits, and are given in Table \ref{tab:differential_number_counts}. The fluxes calculated by this procedure are given in Table \ref{tab:sensitivities}.

\begin{table}
\caption{The values of $\gamma$ used for calculating the confusion limits.}
\begin{tabular}{ |c|c|c|} 
\hline
Telescope & Soft X-ray & Hard X-ray \\
\hline
ATHENA & 1.5 & 1.32 \\
Lynx & 2.22 & 2.29 \\
\hline
\end{tabular}
\label{tab:differential_number_counts}
\end{table}

In Figure \ref{fig:Athena_Lynx_SX}, we show predictions for these two telescopes in the soft X-ray (0.5-2 keV) band. Note that the turnover in the luminosity function seen at low luminosities is due to the halo mass resolution of the dark matter simulation (see Section \ref{sec:lbol_high_z}). As the luminosity limit for Lynx for $z \leq 10$ is below the luminosity of this turnover, the predictions at low luminosities for $z \leq 10$ should be viewed as lower limits on the number densities. This figure also shows how Lynx will be transformational in the study of low luminosity AGNs, and will provide unique constraints and tests of our understanding of black hole physics and galaxy formation. This is a result of increased angular resolution of Lynx compared to ATHENA.

We do not include obscuration for these soft X-ray predictions because at the redshifts we are considering,
the corresponding band in the galaxy rest frame lies at hard X-ray energies - a band for which we are assuming no obscuration.
We show the fiducial model alongside the low accretion efficiency model ($f_{\mathrm{BH}}=0.002$ and $\eta_{\mathrm{Edd}}=16$) and also a model in which the black holes have a seed mass $M_{\mathrm{seed}}=10^5 h^{-1} M_{\odot}$ (compared to the default value $M_{\mathrm{seed}}=10 h^{-1} M_{\odot}$). 

It can be seen how changing the seed black hole mass affects the soft X-ray luminosity function very little at $7 \leq z \leq 9$, and only by a small amount for $L_{SX} < 10^{42} \mathrm{ergs}^{-1}$ at $10<z<15$. This analysis suggests that even high sensitivity telescopes such as Lynx will struggle to differentiate between different seed masses at $7 \leq z \leq 9$ for our model assumptions, but measurements of the number densities of AGNs at low luminosities and very high redshifts ($L_{SX} < 10^{42} \mathrm{ergs}^{-1}$ and $10<z<15$), may be able to exclude models of SMBH seeding that involve high seed masses, although we predict that there will not be a substantial difference in the number densities between these two models.

In Figure \ref{fig:Athena_Lynx_HX} we show the predictions for ATHENA and Lynx in the hard X-ray (2-10 keV) band. For our template SED, an AGN emits
more energy at hard than at soft X-ray energies, but the minimum luminosity of an object that can be detected
is much higher for the hard X-ray band than for the soft X-ray band for ATHENA, while it is only slightly higher for Lynx. This has the effect that for ATHENA, we predcit more AGNs will be detectable in the soft X-ray band compared to the hard X-ray band, whereas for Lynx, we predict that slightly more AGNs will be detectable in the hard X-ray band compared to the soft X-ray band.

For ATHENA, at $z=7$ we predict that $30-80$ AGNs will be detectable per field of view in the soft X-ray band, and $5-20$ for the hard X-ray band (cf. Table \ref{tab:number_of_objects} for the number of objects predicted to be detectable by each survey). At $z=10$, we predict that $0-2$ AGNs will be detectable in the soft X-ray band, and no objects in the hard X-ray band. For Lynx, at $z=7$, we predict that about 800 AGNs per field of view will be detectable in the soft X-ray band, and about $800-900$ in the hard X-ray band. At $z=10$, we predict that about 200 AGNs will be detectable per field of view for both the soft and hard X-ray bands. The low accretion efficiency model predicts fewer AGNs than the fiducial model across all luminosities and redshifts. According to our model, Lynx is the only telescope out of the four studied here that will be able to detect AGNs out to $z=12$, with the possibility of detections at $z=15$, depending on the model variant. More objects are detectable by Lynx compared to ATHENA because Lynx has a better angular resolution, so that it is affected less by source confusion. If we had assumed an ATHENA survey with a 15Ms integration time (as we have assumed for Lynx), the ATHENA survey would not detect any more objects because it is confusion limited for an integration time of 450ks.

\subsection{Properties of detectable AGNs $\&$ SMBHs in high-redshift surveys}
\label{sec:predictions_properties}

\begin{figure*}
\centering
\includegraphics[width=\linewidth]{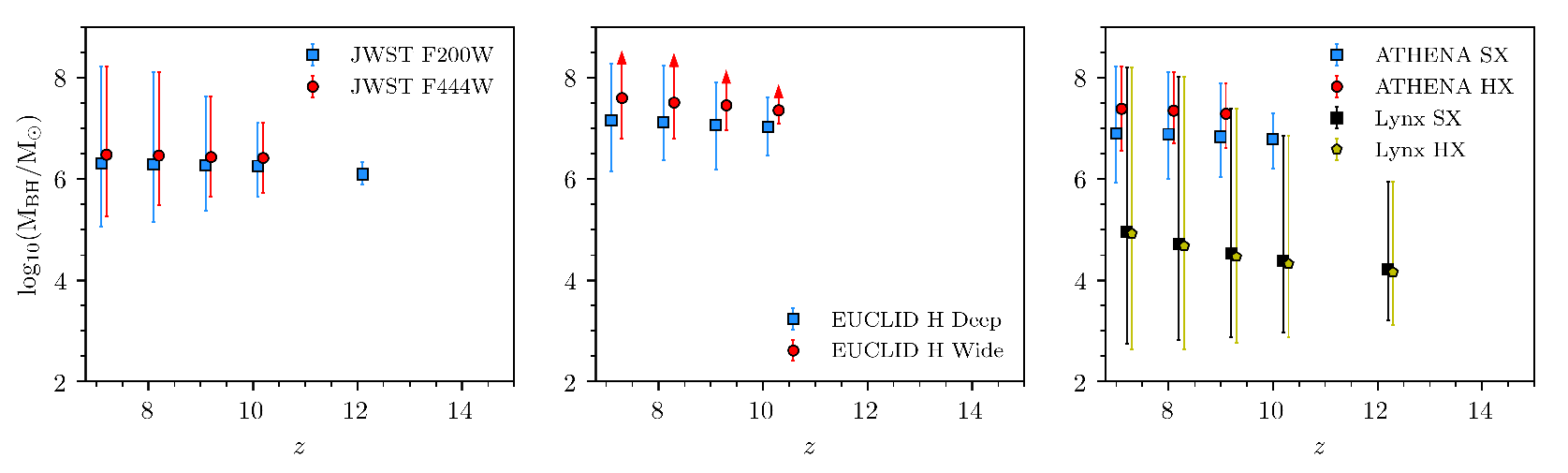}
\caption{The predicted SMBH masses as a function of redshift for AGNs detectable by the surveys with the different telescopes for the fiducial model. Symbols and errorbars show the median and 0-100 percentiles of the distribution of SMBH masses at $z=7,8,9,10,12$. \emph{Left panel:} JWST F200W (blue squares) and JWST F444W (red circles). \emph{Middle panel:} EUCLID H for the Deep survey (blue squares) and for the Wide survey (red squares). The maximum SMBH masses for EUCLID Wide are shown as upward pointing arrows because they are lower limits on the maximum SMBH masses that are detectable. \emph{Right panel:} ATHENA soft and hard X-ray (blue squares and red circles), and Lynx soft and hard X-ray (black squares and green pentagons). In all panels, points for different surveys have been slightly offset in redshift for clarity.}
\label{fig:Nature_objects_evolution_MBH} 
\end{figure*}

\begin{figure*}
\centering
\includegraphics[width=\linewidth]{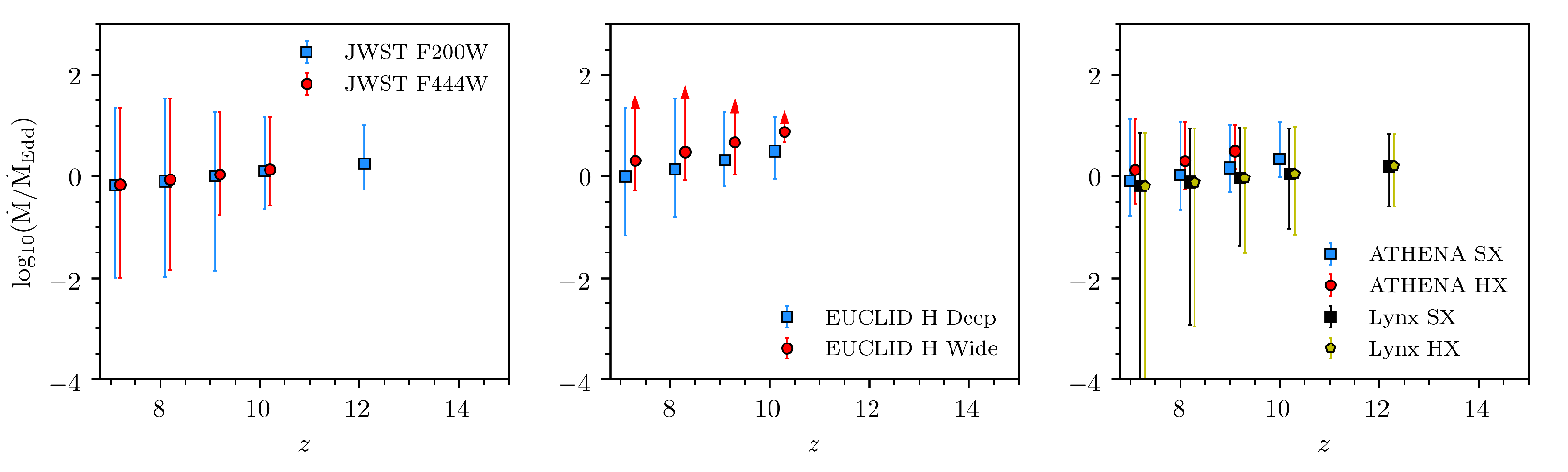}
\caption{The Eddington normalised mass accretion rates as a function of redshift for the AGNs detectable by the surveys with the different telescopes. The lines are as in Figure \ref{fig:Nature_objects_evolution_MBH}.}
\label{fig:Nature_objects_evolution_MDOT} 
\end{figure*}

\begin{figure*}
\centering
\includegraphics[width=\linewidth]{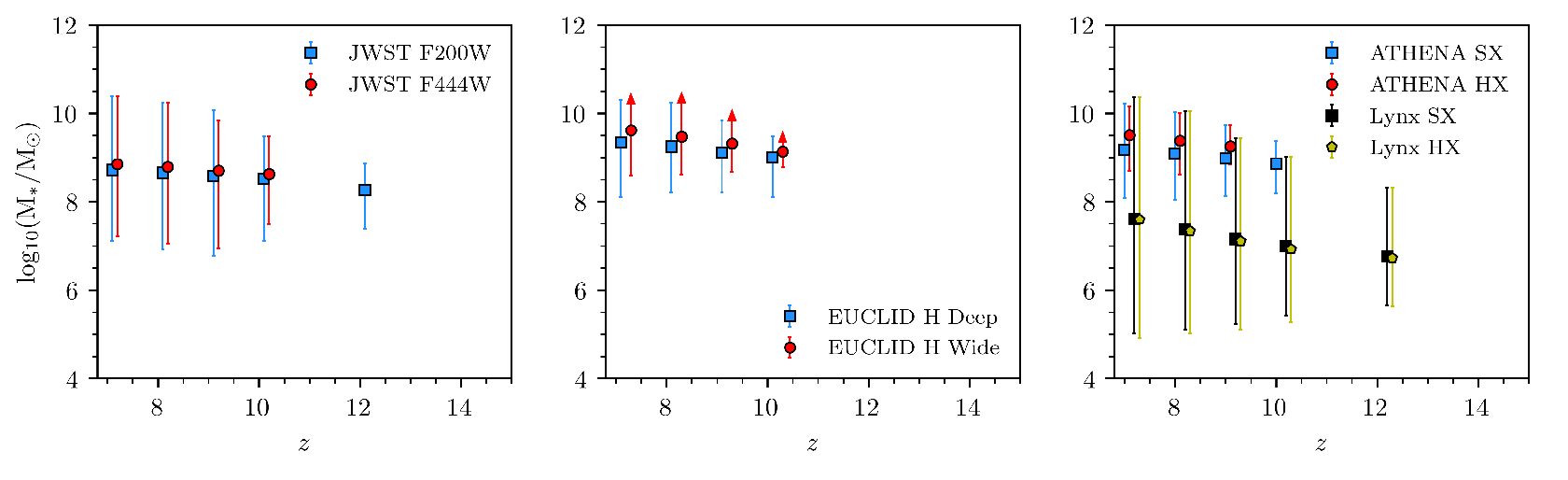}
\caption{The host galaxy stellar masses as a function of redshift for the AGNs detectable by the surveys with the different telescopes. The lines are as in Figure \ref{fig:Nature_objects_evolution_MBH}.}
\label{fig:Nature_objects_evolution_MSTAR} 
\end{figure*}

\begin{figure*}
\centering
\includegraphics[width=\linewidth]{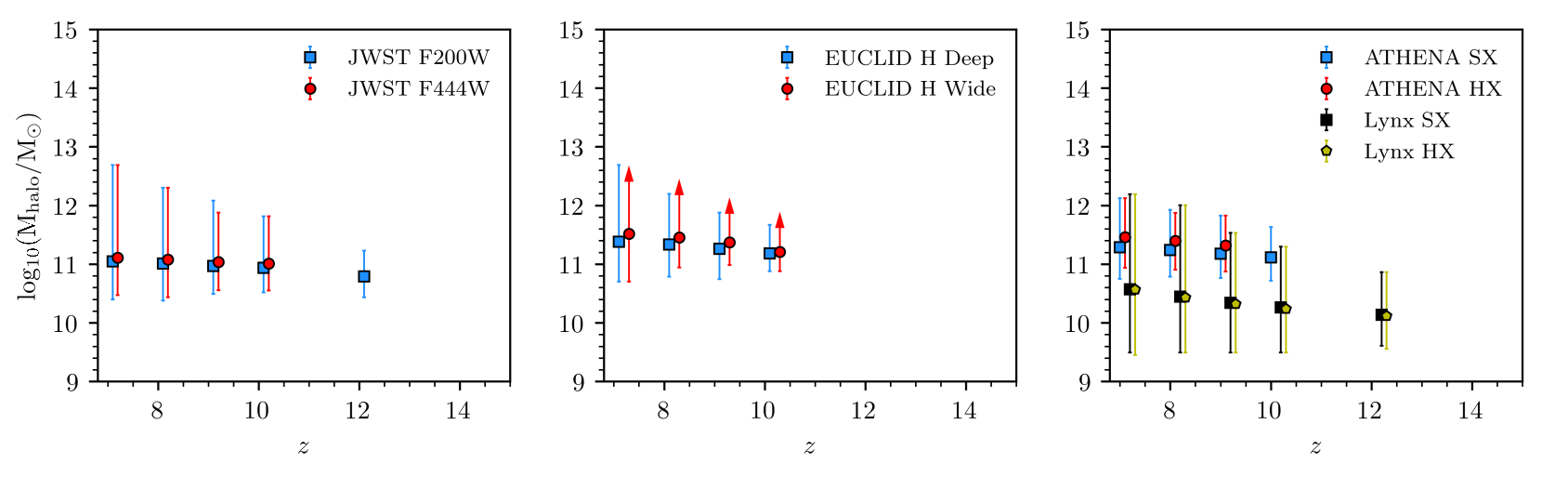}
\caption{The host halo masses as a function of redshift for the AGNs detectable by the surveys with the different telescopes. The lines are as in Figure \ref{fig:Nature_objects_evolution_MBH}. }
\label{fig:Nature_objects_evolution_MHHALO} 
\end{figure*}

We show the predictions for SMBH masses, Eddington normalised mass accretion rates, host galaxy stellar masses, and host halo masses for the AGNs detectable by each survey for redshifts $7 \leq z \leq 15$ in Figures \ref{fig:Nature_objects_evolution_MBH}, \ref{fig:Nature_objects_evolution_MDOT}, \ref{fig:Nature_objects_evolution_MSTAR}, and \ref{fig:Nature_objects_evolution_MHHALO} respectively. We constructed these plots by generating the number density distributions for each property for AGNs above the luminosity limit for the survey at that redshift, and then selecting the part of the distribution with number density above the survey limit, in the same way as we did for luminosity functions in the preceding sections. We then calculated the median, minimum, and maximum values of these distributions, which are plotted in the figures. We also list the median values of these quantities for $z=7$ and $z=10$ in Tables \ref{tab:nature_of_objects_median_z7} and \ref{tab:nature_of_objects_median_z10}. The maximum SMBH masses, Eddington normalised mass accretion rates, galaxy masses, and host halo masses for the EUCLID Wide survey are shown as upward pointing arrows because they are lower limits on the maximum values that EUCLID Wide would detect. This is because the effective survey volume of EUCLID Wide at these redshifts is larger than the volume of the simulation box, and so there may be massive, rare black holes that the survey would detect, but which are not sampled by our simulation volume.

First we compare the near-IR surveys. Compared to EUCLID Deep, we predict that JWST will probe SMBHs with masses about six times lower, in galaxies with stellar masses about four times lower, and in haloes with masses about two times lower, having Eddington normalised accretion rates about 1.4 times lower. 
We predict that the two different EUCLID surveys will detect slightly different populations of AGNs, with EUCLID Wide detecting SMBHs with masses about three times higher, in galaxies with stellar masses about two times higher, and in haloes with masses about 1.4 times higher, having Eddington normalised mass accretion rates about two times higher, compared to EUCLID Deep. 

Now comparing the X-ray surveys, the properties of objects predicted to be detectable in the two ATHENA bands are similar to those predicted to be detectable by EUCLID Deep, but the ATHENA soft X-ray band is predicted to detect SMBHs with masses about two times lower, in galaxies of stellar mass about 1.5 times lower, in host haloes about 1.3 times lower, and having Eddington normalised mass accretion rates about 1.3 times lower, compared to EUCLID Deep. Compared to ATHENA, we predict that Lynx will detect SMBHs with masses about 200 times lower, with galaxy stellar masses about 50 times lower, and in haloes of mass about 10 times lower, with Eddington normalised mass accretion rates about 2 times lower. For each survey, the AGNs detectable at $z=10$ have somewhat lower black hole masses, lower host galaxy stellar masses, lower host halo masses,
and higher Eddington normalised accretion rates than at $z=7$.

Comparing all the distributions of the objects detectable by these surveys at $z=7$, we predict that the objects detectable by the Lynx hard X-ray band will have the lowest median black hole mass, stellar mass, halo mass, and Eddington normalised mass accretion rate. On the other hand, we predict that the obects detectable by the H band in the EUCLID Wide survey will have the highest median black hole mass, stellar mass, halo mass, and Eddington normalised mass accretion rate. 

We predict that Lynx will detect SMBHs that are substantially smaller than in the other surveys, and SMBH host galaxies that are substantially smaller than in the other surveys. Also, Lynx is the only survey that will be able to detect AGNs at $z=7$ in the ADAF accretion state ($\dot{m} < 0.01$). The much lower black hole, galaxy, and halo masses probed by Lynx compared to the other telescopes are a result of it being able to detect AGN at much lower bolometric luminosities. 

While Lynx is predicted here to detect AGNs with smaller black hole masses than the other surveys based on the survey parameters in Table \ref{tab:sensitivities}, we explored whether AGNs with similarly low mass black holes could be detectable by a similarly long integration time with JWST. We considered a 15Ms integration time survey in the JWST F200W band, for a single field of view (compared to our standard assumption of a 10ks integration time in each of 1000 fields of view), assuming the survey is signal-to-noise limited. We predict that for this long integration time survey, JWST could detect objects at $z=7$ down to an AGN bolometric luminosity of $\lbol = 2.8 \times 10^{42} \mathrm{ergs^{-1}}$, compared to $\lbol = 3.8 \times 10^{41} \mathrm{ergs^{-1}}$ for the Lynx soft X-ray band. The smallest black holes at $z=7$ that are detectable by this long integration time JWST survey are of mass $\mbh = 4700 M_{\odot}$, compared to $\mbh = 560 M_{\odot}$ for the Lynx soft X-ray band. JWST is therefore in principle as sensitive as Lynx to low luminosity, low SMBH mass AGNs at high redshift. However, this does not account for the 40 times smaller field of view of JWST compared to Lynx, which greatly reduces the survey volume, nor the greater difficulty of separating the light of the AGN from that of the host galaxy in near-IR compared to X-rays. 

The largest detectable SMBH is also different for each of these surveys. Surveys with larger survey areas can probe down to lower number densities, and so generally can detect higher mass SMBHs. However, because the black hole mass function decreases fairly steeply at the high mass end, increasing the survey area only slightly increases the mass of the largest SMBH detectable. For halo masses, a larger survey area does not necessarily correspond to detecting larger haloes from the AGNs they contain, because the largest haloes can host lower luminosity objects (see Figure \ref{fig:Lbol_Mhalo_z7_z10}). Therefore the maximum halo mass is also affected by the sensitivity limit, as seen for ATHENA and Lynx in the right panel of Figure \ref{fig:Nature_objects_evolution_MHHALO}. A similar argument can be applied for stellar masses as seen in Figure \ref{fig:Nature_objects_evolution_MSTAR}. 

We also explored the effect of halo mass resolution in our simulation on the properties of objects detectable by these surveys (see Section \ref{sec:lbol_high_z}). We find that if we degrade the halo mass resolution, as long as the objects have bolometric luminosities above the value at which the luminosity functions converge (i.e. $\lbol > 10^{43} \mathrm{ergs^{-1}}$), the properties of the black holes are the same. The predictions of black hole properties for surveys by JWST, EUCLID and ATHENA are insensitive to this effect, but for Lynx the values given should be regarded as upper limits.


\section{Conclusions}
\label{sec:conclusions}

\vfill

Recent advances in observational capabilities have opened up studies of the high-redshift Universe, but many uncertainties regarding the early stages of galaxy formation and evolution remain. The origin of supermassive black holes (SMBHs) and their role in the early Universe still remains a mystery. Fortunately the next decade-and-a-half offers us exciting new opportunities to probe the high redshift Universe,
especially given the plans for powerful new space-based telescopes such as JWST and EUCLID at optical/near-IR wavelengths, and ATHENA and Lynx at X-ray energies. These will
offer us a multiwavelength view of the distant Universe and allow us to characterise 
physical processes in galaxy formation. The role of SMBHs and their growth in the distant Universe will be  
probed with much greater accuracy than ever before.

With these potential new developments in mind, we present predictions for AGNs in the high redshift Universe ($z \geq 7$) using the semi-analytic model of galaxy formation \galform. In \galform, galaxies (and hence AGNs) form in dark matter haloes, with the evolution of the dark matter haloes described by halo merger trees. Here, the merger trees have been generated from a dark matter N-body simulation. In the model, SMBHs grow by accretion of gas during starbursts triggered either by mergers or disc instabilities, or by accretion of gas from hot gas halos, or by merging with other SMBHs. The evolution of the SMBH spin is also calculated in the model with SMBHs changing spin either by accretion of gas, or by merging with another SMBH. From the SMBH mass accretion rates, AGN bolometric luminosities are then calculated, which when combined with empirical SED and obscuration models can be used to calculate luminosities in different bands. The \galform model used here is that presented in \cite{griffin19}, which showed that the predicted AGN luminosity functions are in good agreement with observational data at $0\leq z \leq 5$. 

We present model predictions for the AGN bolometric luminosity function for $7 \leq z \leq 15$, finding that it evolves to lower luminosities and lower number densities at higher redshift as a result of hierarchical structure formation. When we split the bolometric luminosity function at these redshifts by accretion disc mode and gas fuelling mode, we find that the dominant accretion disc modes
are thin discs at low luminosities ($\lbol < 10^{45} \mathrm{ergs^{-1}}$), and super-Eddington objects at higher luminosities, and the dominant 
gas fuelling mode at all luminosities is starbursts triggered by disc instabilities. The model allows SMBHs to grow at mass accretion rates above the Eddington rate, so when we limit the SMBH gas accretion rate to the Eddington rate, the number of SMBHs at high redshift is significantly reduced. We also explore the effect of varying the SMBH seed mass on the bolometric luminosity function. We find that when we use a much larger seed black hole mass ($10^5 h^{-1} M_{\odot}$ compared to $10h^{-1} M_{\odot}$ in the fiducial model), the luminosity functions are relatively unaffected, except for $\lbol < 10^{43} \mathrm{ergs^{-1}}$ for $z>10$. 

We then present predictions for JWST, EUCLID, 
ATHENA, and Lynx, using sensitivities and survey areas for possible surveys with these telescopes. For example, we assume a $1.5 \times 10^{7}$s exposure for Lynx over a survey area of 360 arcmin$^2$ (1 field of view), whereas we assume a thousand $10^{4}$s exposures for JWST over a total survey area of 9680 arcmin$^2$ (1000 fields of view). We find that the different surveys will probe down to different AGN bolometric luminosities and number densities, and hence sample different parts of the AGN population. 

We also present predictions for two variants to the fiducial model that provide a better fit to the rest-frame UV and rest-frame soft X-ray luminosity 
functions of AGNs at $z=6$. In these models we vary either the amount of AGN obscuration or the SMBH accretion efficiency (defined here as the fraction of gas accreted onto the SMBH in a starburst). The resulting luminosity functions have lower number densities by factors of about 4 and 2 respectively. AGN obscuration and SMBH accretion efficiency are both uncertainties for the AGN population at high redshift. Comparing these predictions to observations should allow 
us to better both of these aspects at high redshift.

The properties of the SMBHs and AGNs detectable depend on the survey and wavelength. For our fiducial model, we predict that the AGNs detectable at $z=7$ will have median black hole masses that vary from $8 \times 10^4 M_{\odot}$ to $4 \times 10^7 M_{\odot}$, and median Eddington normalised mass accretion rates that vary from 
$0.6-2$. These AGNs are predicted to reside in host galaxies with median stellar masses that vary from $4 \times 10^7 M_{\odot}$ to $4 \times 10^9 M_{\odot}$, and in haloes with median masses from $4 \times 10^{10} M_{\odot}$ to $3 \times 10^{11} M_{\odot}$. At $z=10$, the AGNs detectable are predicted to have black hole masses that vary between $2 \times 10^4 M_{\odot}$ to $2 \times 10^7 M_{\odot}$, with Eddington normalised mass accretion rates that vary from 
$1-8$. The host galaxies of these AGNs are predicted to have masses that vary from $8 \times 10^6 M_{\odot}$ to $1 \times 10^9 M_{\odot}$, in haloes with masses that very from $2 \times 10^{10} M_{\odot}$ to $2 \times 10^{11} M_{\odot}$. The different telescopes will therefore provide different but complementary views on the $z>6$ AGN population. For the survey parameters assumed here, Lynx is predicted to detect SMBHs with the lowest masses, in the lowest mass host galaxies and lowest mass host haloes, and so will provide the best opportunity to probe the nature of SMBH seeds. However, a similarly long integration (15Ms) in a single field of view with JWST could in principle detect similarly faint AGN at high redshift. 

These future telescopes should therefore be able to detect SMBHs at very high redshift having masses $\sim 10^{4}-10^{5} M_{\odot}$ that are comparable to those of the highest mass seed SMBHs that are envisaged in current scenarios, and put improved constraints on the physical mechanisms by which these seed SMBHs form.


\section*{Acknowledgements}

We thank Steve Warren for helpful comments on an earlier version of this paper. This work was supported by the Science and Technology facilities Council grants ST/L00075X/1 and ST/P000541/1.
AJG acknowledges an STFC studentship funded by STFC grant ST/N50404X/1. 
CDL has received funding from the ARC Centre of Excellence for All Sky Astrophysics in 3 Dimensions (ASTRO 3D), through project number CE170100013. CDL also thanks the MERAC Foundation for a Postdoctoral Research Award.
This work used the DiRAC Data Centric system at Durham University, operated by
the Institute for Computational Cosmology on behalf of the STFC DiRAC HPC
Facility (www.dirac.ac.uk). This equipment was funded by BIS National
E-infrastructure capital grant ST/K00042X/1, STFC capital grants ST/H008519/1
and ST/K00087X/1, STFC DiRAC Operations grant ST/K003267/1 and Durham
University. DiRAC is part of the National E-Infrastructure.



\bibliographystyle{mnras}
\bibliography{references}





\appendix

\section{Effect of halo mass resolution}
\label{app:resolution_effects}

\begin{figure}
\centering
\includegraphics[width=\linewidth]{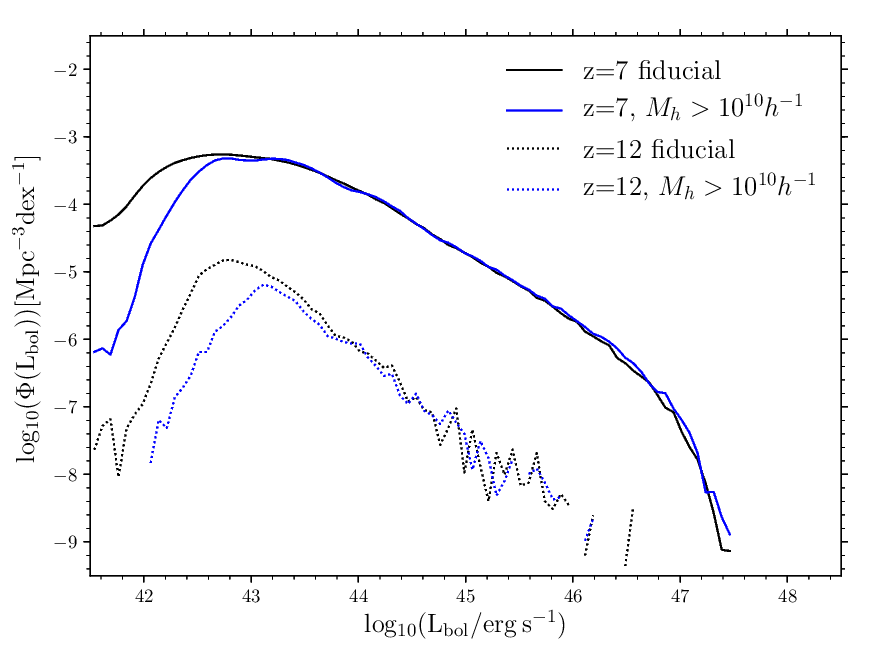}
\caption{The bolometric luminosity function at $z=7$ (solid lines), and $z=12$ (dotted lines) for the halo mass resolution of $2.12 \times 10^{9} h^{-1} M_{\odot}$ as for the standard model (black lines) and for a halo mass resolution of $10^{10} h^{-1} M_{\odot}$ (blue lines).}
\label{fig:Resolution_bol_lf} 
\end{figure}

In Figure \ref{fig:Resolution_bol_lf} we show the predicted bolometric luminosity function at $z=7$ and $z=12$ for the fiducial model, which has a halo mass resolution of $2.12 \times 10^{9} h^{-1} M_{\odot}$, and for a halo mass resolution of $10^{10}h^{-1} M_{\odot}$. The figure demonstrates that the turnover seen in the luminosity function at $\lbol \sim 10^{43} \mathrm{ergs}^{-1}$ is due to the dark matter simulation only resolving haloes above a certain mass. The two bolometric luminosity functions are converged for $\lbol \gtrsim 10^{43} \mathrm{ergs}^{-1}$ (depending somewhat on redshift), while the poorer halo mass resolution leads to fewer objects for $\lbol < 10^{43} \mathrm{ergs}^{-1}$.

\section{The effect of the SMBH seed mass}
\label{app:seed_mass}

\begin{figure}
\centering
\includegraphics[width=\linewidth]{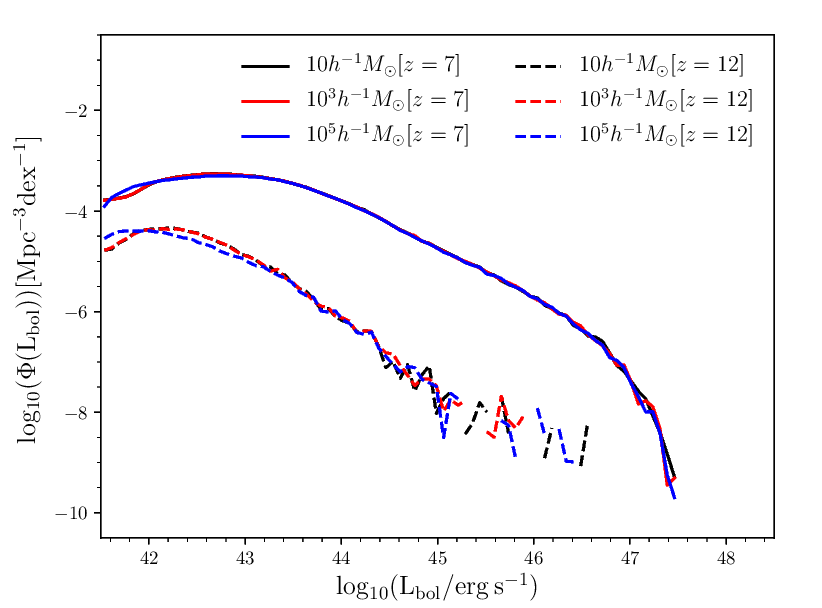}
\caption{The bolometric luminosity function at $z=7$ (solid lines), and $z=12$ (dashed lines) for seed masses of $10h^{-1} M_{\odot}$ (black),
$10^{3}h^{-1} M_{\odot}$ (red) and $10^{5}h^{-1} M_{\odot}$ (blue). Note that the black lines are underneath the red lines.}
\label{fig:Seed_bol} 
\end{figure}

In Figure \ref{fig:Seed_bol} we show the AGN bolometric luminosity function at $z=7$ and $z=12$ for three different seed masses ($10h^{-1} M_{\odot}$, $10^{3}h^{-1} M_{\odot}$, and $10^{5}h^{-1} M_{\odot}$).
The luminosity functions for the three different seed masses are consistent with each other within statistical errors for $\lbol > 10^{42}$ ergs$^{-1}$ at $z=7$, and consistent with each other for $\lbol > 10^{43}$ ergs$^{-1}$ at $z=10$.

\section{Number of detectable objects}

In Table \ref{tab:number_of_objects} we show the number of objects detectable by each survey at $z=7$, $z=9$, $z=10$, and $z=12$, with sensitivities and survey areas as in Table \ref{tab:sensitivities}.

\begin{table*}
\caption{Predictions for the number of AGNs expected to be detectable at different redshifts by the different telescopes, using the sensitivity limits and survey areas given in Table \ref{tab:sensitivities}. The ranges of values correspond to the three different variants of the model (see Section \ref{sec:variants}): the fiducial model, which uses the LZMH obscuration fraction, the fiducial model using the Z6MH obscuration fraction, and the low accretion efficiency model.}
\begin{tabular}{ |c|c|c|c|c|c| } 
\hline
Instrument & Filter & $z=7$ & $z=9$ & $z=10$ & $z=12$ \\ 
\hline
JWST & F200W & 90-500 & 5-30 & 1-8 & 0 \\
 & F444W & 60-300 & 3-20 & 0-4 & 0 \\
\hline
EUCLID Deep & H & 100-600 & 5-20 & 1-5 & 0 \\
\hline
EUCLID Wide & H & 8000-30000 & 300-1000 & 70-300 & 1-4 \\
\hline
ATHENA WFI & Soft X-ray & 30-80 & 1-4 & 0-2 & 0-1 \\
& Hard X-ray & 5-20 & 0 & 0 & 0 \\
\hline
Lynx & Soft X-ray & 800 & 200-300 & 200 & 100-200 \\
& Hard X-ray & 800-900 & 200-300 & 200 & 100-200 \\
\hline
\end{tabular}
\label{tab:number_of_objects}
\end{table*}

\section{Properties of detectable objects}

In Tables \ref{tab:nature_of_objects_median_z7} and \ref{tab:nature_of_objects_median_z10} we show the median SMBH masses, Eddington normalised accretion rates, host galaxy stellar masses and host halo masses of AGNs detectable by the future surveys at $z=7$ and $z=10$. The assumed sensitivities and survey areas are given in Table \ref{tab:sensitivities}.

\begin{table*}
\caption{The median SMBH masses, Eddington normalised mass accretion rates, host galaxy stellar masses, and host halo masses of the AGNs predicted to be detectable by JWST, EUCLID, ATHENA, and Lynx at $z=7$ for our fiducial model, for the survey parameters given in Table \ref{tab:sensitivities}.}
\begin{tabular}{ |c|c|c|c|c|c| } 
\hline
Instrument & Filter & $M_{\mathrm{SMBH}} (M_{\odot})$ & $\dot{m} = \dot{M}/ \dot{M}_{\mathrm{Edd}}$ & $M_{\star} (M_{\odot})$ & $M_{\mathrm{halo}} (M_{\odot})$ \\ 
\hline
JWST & F200W & $2.0 \times 10^6$ & $0.7$ & $5.2 \times 10^8$ & $1.1 \times 10^{11}$ \\
 & F444W & $3.0 \times 10^6$ & $0.7$ & $7.1 \times 10^8$ & $1.3 \times 10^{11}$\\
\hline
EUCLID Deep & H & $1.4 \times 10^7$ & $1.0$ & $2.2 \times 10^9$ & $2.4 \times 10^{11}$\\
\hline
EUCLID Wide & H & $4.0 \times 10^7$ & $2.0$ & $4.1 \times 10^9$ & $3.3 \times 10^{11}$ \\
\hline
ATHENA WFI & Soft X-ray & $8.0 \times 10^6$ & $0.8$ & $1.5 \times 10^9$ & $1.9 \times 10^{11}$ \\
          & Hard X-ray & $2.4 \times 10^7$ & $1.3$ & $3.2 \times 10^9$ & $2.9 \times 10^{11}$ \\
\hline
Lynx & Soft X-ray & $8.9 \times 10^4$ & $0.6$ & $4.1 \times 10^7$ & $3.7 \times 10^{10}$ \\
          & Hard X-ray & $8.2 \times 10^4$ & $0.6$ & $3.9 \times 10^7$ & $3.6 \times 10^{10}$\\
\hline
\end{tabular}
\label{tab:nature_of_objects_median_z7}
\end{table*}

\begin{table*}
\caption{The same as Table \ref{tab:nature_of_objects_median_z7}, but at $z=10$. We predict that the ATHENA hard X-ray band will not be able to detect any AGNs at $z=10$.}
\begin{tabular}{ |c|c|c|c|c|c| } 
\hline
Instrument & Filter & $M_{\mathrm{SMBH}} (M_{\odot})$ & $\dot{m} = \dot{M}/ \dot{M}_{\mathrm{Edd}}$ & $M_{\star} (M_{\odot})$ & $M_{\mathrm{halo}} (M_{\odot})$ \\ 
\hline
JWST & F200W & $1.8 \times 10^6$ & $1.2$ & $3.2 \times 10^8$ & $8.6 \times 10^{10}$\\
 & F444W & $2.6 \times 10^6$ & $1.4$ & $4.2 \times 10^8$ & $1.1 \times 10^{11}$ \\
\hline
EUCLID Deep & H & $1.1 \times 10^7$ & $3.2$ & $1.0 \times 10^9$ & $1.5 \times 10^{11}$ \\
\hline
EUCLID Wide & H & $2.2 \times 10^7$ & $7.5$ & $1.4 \times 10^9$ & $1.6 \times 10^{11}$\\
\hline
ATHENA WFI & Soft X-ray & $6.0 \times 10^6$ & $2.1$ & $7.3 \times 10^8$ & $1.3 \times 10^{11}$\\
          & Hard X-ray & - & - & - & - \\
\hline
Lynx & Soft X-ray & $2.4 \times 10^4$ & $1.1$ & $9.8 \times 10^6$ & $1.8 \times 10^{10}$ \\
          & Hard X-ray & $2.1 \times 10^4$ & $1.1$ & $8.4 \times 10^6$ & $1.7 \times 10^{10}$ \\
\hline
\end{tabular}
\label{tab:nature_of_objects_median_z10}
\end{table*}

\bsp	
\label{lastpage}
\end{document}